\documentclass[conference]{IEEEtran}
\IEEEoverridecommandlockouts
\usepackage{cite}
\usepackage{amsmath,amssymb,amsfonts}
\usepackage{algorithmic}
\usepackage{graphicx, caption, subcaption}
\usepackage{tikz}
\usepackage{textcomp}
\usepackage{xcolor, soul}
\usepackage{comment}
\usepackage{multicol}
\usepackage{multirow}
\usepackage{booktabs}
\usepackage{tablefootnote,tabularx}

\newcommand*\circled[1]{\tikz[baseline=(char.base)]{
            \node[shape=circle,fill,inner sep=2pt] (char) {\textcolor{white}{#1}};}}

\def\BibTeX{{\rm B\kern-.05em{\sc i\kern-.025em b}\kern-.08em
    T\kern-.1667em\lower.7ex\hbox{E}\kern-.125emX}}
\begin{document}

\title{Using Undervolting as an on-Device Defense Against Adversarial Machine Learning Attacks\thanks{This work was funded in part by the National Science Foundation under awards CCF-2028944 and CCF-1629392 and by the Air Force Research Laboratory under the Assured and Trusted Microelectronics Solutions award FA8650-20-C-1719.}}

\author{
    \IEEEauthorblockN{Saikat Majumdar, Mohammad Hossein Samavatian, Kristin Barber, Radu Teodorescu}
    \IEEEauthorblockA{Department of Computer Science and Engineering \\ The Ohio State University, Columbus, OH, USA
    \\ \{majumdar.42, samavatian.1, barber.245, teodorescu.1\}@osu.edu}
}



\maketitle
\thispagestyle{plain}
\pagestyle{plain}

\begin{abstract}
Deep neural network (DNN) classifiers are powerful tools that drive a broad spectrum of important applications, from image recognition to autonomous vehicles. Unfortunately, DNNs are known to be vulnerable to adversarial attacks that affect virtually all state-of-the-art models. These attacks make small imperceptible modifications to inputs that are sufficient to induce the DNNs to produce the wrong classification. 

In this paper we propose a novel, lightweight adversarial correction and/or detection mechanism for image classifiers that relies on undervolting (running a chip at a voltage that is slightly below its safe margin). We propose using controlled undervolting of the chip running the inference process in order to introduce a limited number of compute errors. We show that these errors disrupt the adversarial input in a way that can be used either to correct the classification or detect the input as adversarial. We evaluate the proposed solution in an FPGA design and through software simulation. We evaluate 10 attacks and show average detection rates of 77\% and 90\% on two popular DNNs. 

\end{abstract}


\section{Introduction}
\label{sec:introduction}

Deep neural networks (DNNs) are emerging as the backbone of an increasing number of diverse applications. Some of these applications benefit from the deployment of sophisticated DNN models into so-called edge devices such as smartphones, smart home devices, autonomous driving systems, healthcare and many others \cite{bojarski2016end, litjens2017survey, mohammadi2018deep}. Some of these applications, such as autonomous driving, are safety critical and their failure can endanger lives. Unfortunately, DNNs are known to be vulnerable to a variety of security threats. One of these threats is the so-called ``adversarial attack'' against DNN classifiers. The goal of these attacks is to induce the DNN to misclassify an input that the attacker controls, into the wrong class. This is achieved by slightly altering the input to the classifier model in a way that induces it to produce erroneous an erroneous result. For example, the image of a {\em STOP} sign can be slightly modified to cause a road sign recognition model to classify it as a {\em YIELD} sign in spite of those changes being imperceptible to the human eye. 

A broad spectrum of prior work has demonstrated successful attacks on image classifiers and other computer vision applications \cite{carlini2017towards, moosavi-dezfooli2017universal, moosavi-dezfooli2016deepfool, goodfellow2015explaining, chen2018ead, kurakin2017adversarial, papernot2016the, madry2018towards}. Prior work has also shown that virtually all classifiers are vulnerable including the popular ResNet\cite{ResNet}, AlexNet\cite{NIPS2012_4824}, VGG\cite{simonyan2015deep}. The most recent, strongest attacks \cite{carlini2017towards, chen2018ead} achieve reliable misclassification with changes to the input images that are small enough to be undetectable by casual observation. 

In response to these attacks, prior work has proposed a suite a defenses \cite{ma2019nic,papernot2016distillation,xu2018feature,dhillon2018stochastic,cao2017mitigating,ma2018characterizing}. These defenses generally rely either on model retraining, which are not easily generalizable, or online, inference-time defenses, which are mostly software-based and have very high overheads. Only a few hardware-accelerated defenses have been proposed to date. DNNGuard \cite{wang2020dnnguard} is one example that relies on a separate dedicated classifier for adversarial detection, which also comes at a high cost. 

Prior work such as \cite{hu2019a} has shown that adding some amount of random noise to inputs can help DNNs correctly classify adversarial inputs. Recent work by Guesmi et al.\ \cite{guesmi2020defensive} proposed using approximate computation to introduce a controlled number of error in the DNN inference to correct some of the adversarial input into valid classifications. The observation that these and other prior work has made is that some amount of alteration to the model or input tends to affect adversarial inputs more than benign ones. 

In this paper we propose a new, lightweight adversarial correction and/or detection mechanism for image classifiers that relies on the compute errors introduced by undervolting (running a chip at a voltage that is slightly below its safe margin). Prior work has shown that DNNs tend to be resilient to the errors introduced by undervolting \cite{salami2020experimental, 8574581} in an FPGA. In this work we perform a thorough characterization of the effects of undervolting errors on classification of both adversarial and benign inputs. We show that errors during inference lead classifiers to assign a different class to adversarial images than would be assigned in the error-free inference. The new classification is either the correct class for the input (before the attack) or is a third class that is neither the original or the intended target of the attacker. We also show that undervolting is much less likely to lead to the misclassification of benign (unaltered) images. Based on these empirical observations, we propose a lightweight, energy efficient defense mechanism targeted at deployment on energy constrained edge devices. 

Our proposed solution uses undervolting to achieve two goals: (1) introduce random errors caused by low-voltage operation into the DNN inference execution and (2) save energy by operating at the lower voltage. 

Prior work has shown that errors introduced by undervolting are both random (they occur in random locations in different chips) and repeatable (they occur in the same location within a chip) \cite{bacha-isca13, 9152636}. The random distribution makes the defense unique to each chip, making it harder for an attacker to generate a successful general attack. The repeatably of the errors allows for system control over the approximate error rate that a chip is likely to encounter, allowing chips to continue execution without system crashes, in the presence of occasional localized errors. Prior work has used such controlled errors in security, for example, as a mechanism for hardware-backed authentication \cite{bacha-micro15} or to attack SGX secure enclaves in Intel processors \cite{9152636}. We propose using the same controlled errors to inject noise in the processor while running a DNN classifier. 

We investigate two possible designs: one focused on adversarial correction, the other focused on detection. In the correction-based mechanism the system lowers supply voltage to a pre-determined set point, known to induce a certain number of errors, while the system is running the inference process. We show that this very simple approach leads the classifier to correctly classify an average of 50-55\% of adversarial inputs, generated with multiple attacks. The second mechanism, focused on adversarial detection, compares the classification outcome of an error-free reference run with the undervolted run. If they are different, the input is classified as adversarial. Detection rate averages across a range of attacks are 77\% and 90\% on DenseNet and VGG16, respectively. 

We also developed a lightweight on-device training process that fine tunes the model to the error profile on each device. This process improves the classification accuracy for benign (non-adversarial) inputs in the presence of undervolting errors, while keeping the correction/detection rates of adversarials largely unchanged. Finally, we show that the proposed defense is resilient against attacks that have knowledge of the defense and attempt to circumvent it by training on a model with random errors or a fixed error distribution. 

We evaluate the proposed system with an FPGA implementation of a DNN accelerator (which we undervolt with an external controller) as well as an error-model based software implementation that allows us to measure the impact of multiple error distributions and rates. We evaluate 10 attacks or attack variants on two popular deep convolutional neural networks, VGG16\cite{simonyan2015deep} and DenseNet\cite{huang2018densely}, on the ImageNet\cite{krizhevsky2012imagenet} and CIFAR\cite{cifar10} datasets. 

This paper makes the following contributions: 
\begin{itemize}
    \item The first work we are aware of to propose using undervolting errors as a defense against ML adversarial attacks. 
    \item A lightweight on-device training mechanism to improve accuracy of benign inputs under errors. 
    \item A thorough study on the effects of different error rates and distributions on the adversarial detection/correction rates as well as the effects on accuracy of benign classification. 
    \item Evaluates the proposed system on a FPGA platform and using a software-based model.  
\end{itemize}

The rest of this paper is organized as follows: Section \ref{sec:related} presents background and related work. Section \ref{sec:threat} details the threat model. Section \ref{sec:design} presents the details of the proposed defenses. Sections \ref{sec:method} and \ref{sec:eval} present the evaluation and Section \ref{sec:conclusion} concludes. 




\section{Background and Related Work}
\label{sec:related}

\subsection{Undervolting}
    

Voltage underscaling is a common power saving approach, whereby reducing the supply voltage causes a significant reduction in power consumption. A large body of work \cite{razor,lefurgy-micro12,bacha-isca13} has explored approaches to dynamically reduce voltage margins to current operating conditions, in a process generally referred to as "undervolting". For example, the well-known Razor~\cite{razor} design dynamically lowers supply voltage until occasional timing errors occur. Additional latches running on a delayed clock are used on the vulnerable paths to detect and recover from these errors. Other work by Lefurgy et al.\ \cite{lefurgy-micro12} uses hardware critical path monitors specially built into the chip to detect when the processor approaches its timing margin as a result of undervolting. Salami et al. \cite{salami2020experimental, 8574581}  have characterized extensively the behavior of FPGAs running a DNN accelerator with undervolting. They also provide experimental analysis of undervolting below the safe voltage, which leads to observable faults that manifest as bit flips. Similar studies have been performed by undervolting various system components in CPUs \cite{bacha-isca13}, \cite{8686519}, \cite{8270368}, \cite{7011402}, GPUs \cite{7056030}, DRAMs \cite{inproceedings} focusing primarily on fault characterization, voltage guardband analysis and fault mitigation. 

\subsection{Adversarial Machine Learning Attacks}

Szegedy et al. in \cite{szegedy2014intriguing} showed that small perturbations to the inputs can force machine learning models to misclassify. 
Most attacks follow the same general approach. Let $f$ be a probabilistic classifier with logits $f_y$ and let $F(x) = \arg\max_yf_y(x)$. The goal of the adversary is to find an $L^p$-norm bounded perturbation such that $\Delta x \in B^p_\epsilon(0):=\{\Delta:||\Delta||_p \leq \epsilon\}$, where $\epsilon$ controls the attack strength, such that the perturbed sample $x+\Delta x$ gets misclassified by the classifier $F(x)$. The attacker's goal can be formulated as:
\begin{equation} \label{eq1}
\text{given}~x, \text{find}~x'
~~\text{s.t.}~||x'-x||_p \leq \epsilon_{max} ~\text{and}~ F(x') \neq y
\end{equation}

In the case of \emph{targeted} attacks, the adversary's goal is for \textbf{x$^\prime$} to be misclassified as $F(x')=t$ with $t$ being a class different than $y$.

\textbf{FGSM} \cite{goodfellow2015explaining} is a fast, simple and efficient method for generating adversarial attacks using the $L_{\infty}$ norm that measures the gradient of the loss with respect to the input data, then adjusts the input data to maximize the loss.
The {\bf Carlini-Wagner (CW)} attack \cite{carlini2017towards} finds adversarial examples with considerably smaller perturbations and higher accuracy. CW has variants for the $L_0$ and $L_\infty$ distortion metrics. 
 CW-$L_0$ uses the CW-$L_2$ attack to identify pixels which do not contribute much to the classifier output and freezes those pixel values. CW-$L_\infty$ revises the objective function to limit perturbations to be less than a threshold, iteratively searching for the smallest possible threshold.
The \textbf{EAD} attack \cite{chen2018ead} formulates the objective function in terms of regularized elastic-nets, incorporating the $L_1$ distance metric for distortion
, where elastic-net regularization is a linear mixture of $L_1$ and $L_2$ penalty functions.
EAD has two variants, where the optimization process can be tuned by two different decision rules: \emph{EN}, the elastic-net loss or \emph{$L_1$}, the least $L_1$ distortion.
\textbf{DeepFool} \cite{moosavi-dezfooli2016deepfool} proposed a method to generate adversarial inputs by searching for the closest distance from the benign image to the decision boundary of the target image. We test our defense against multiple variants of the aforementioned attacks. 


\subsection{Defensive Techniques}

Current defense methods against adversarial attacks can be divided into the following broad categories:

1. Making the network itself robust \cite{WardeFarley20161AP},\cite{papernot2016science},\cite{madry2018towards}, \cite{Kurakin2017AdversarialML} by hardening the model, for example, by using \textit{Adversarial Training}. The training set can be augmented with adversarial samples so that the network learns to classify adversarial samples correctly. In general, this class of defenses relies on known attacks and is difficult to generalize \cite{papernot2016science}. 

2. Transforming the input samples \cite{xu2018feature, song2018pixeldefend, liao2018defense} which pre-processes the model inputs, such as by encoding or filtering, to denoise the adversarial perturbation. 

Prior work has also attempted to detect adversarial inputs. Approaches proposed by \cite{xu2018feature} aim to ``squeeze" an input image and \cite{dhillon2018stochastic} applies weighted dropouts to neurons, to detect an adversarial sample. Grosse et al. \cite{grosse2017statistical} proposed a statistical test using maximum mean discrepancy and suggests the energy distance as the statistical distance measure for detecting adversarials. Other approaches like \cite{metzen2017detecting, grosse2017statistical} suggest training a detector using adversarial examples. 

In general, methods that rely on re-training are not generalizable. Detection approaches are more successful, but their overhead is generally very high. For instance, Feature Squeezing \cite{xu2018feature} has a 3-4$\times$ overhead because it relies on 4 models for detection. Our proposed solution is lightweight and highly effective at detecting a wide variety of adversarial attacks.

\section{Threat Model}
\label{sec:threat}

This paper assumes adversaries that have complete access to the DNN model, with full knowledge of the architecture and parameters, and are able train their attacks accordingly. We assume an adversary that is able to directly deliver inputs to the DNN classifier, such as in \cite{eykholt2018robust,pautov2019adversarial,raju2020blurnet} where attackers generate visual  adversarial perturbations that are supposed to mimic real-world obstacles or create adversarial traffic signs to attack autonomous driving. We focus mainly on recent state-of-the-art optimization-based attacks--CW and EAD--since it has been demonstrated that all earlier attacks can be overcome utilizing other methods, such as adversarial training \cite{goodfellow2015explaining} or defensive distillation \cite{papernot2016distillation}, which could be used in combination with our approach. Finally, our defense is optimized for edge devices such as smartphones or autonomous vehicles, rather than server or cloud-based classifiers. 
\section{Defense Design}
\label{sec:design}

    


Lowering the supply voltage of chips leads to significant power savings. However, reduced voltage operation may also introduce some faults in the device. This is because propagation latency increases at lower supply voltages, which can lead to timing faults. These timing faults can manifest as \textit{bit-flips} in logic outputs or memory. DNNs are known to be resilient to some errors, generally experiencing a graceful degradation in accuracy. 


\subsection{Using Undervolting as a Defense Mechanism}


It has been shown \cite{cohen2019certified, he2019parametric, lecuyer2019certified} that  adding  some amount of random noise to images can help DNNs correctly detect adversarial inputs. 
The key observation we make in this work is that injecting noise directly in the model improves the ability to detect stronger adversarial inputs, for which correct classification cannot be achieved. 

\begin{figure}[htbp]
\centering
     \includegraphics[width=0.6\linewidth]{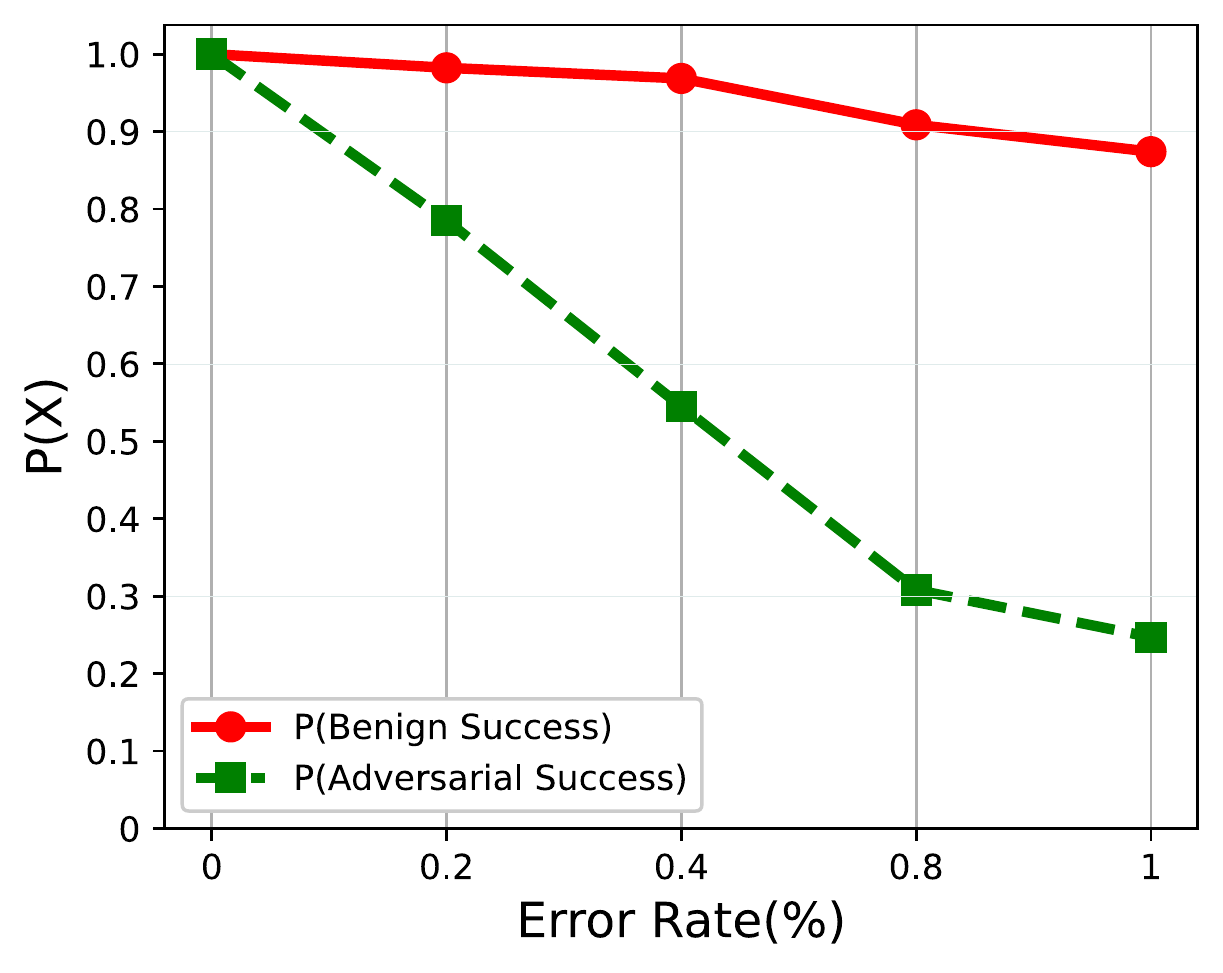}
  \caption{\small Probability of success for benign and adversarial classification under the effects of random errors in a DNN model.}\label{fig:benvsadv}
\end{figure}

Figure \ref{fig:benvsadv} shows the response of a DNN classifier to increasing levels of random noise injected into the model, while classifying both adversarial and benign inputs. We can see that, as the error rate increases the adversarial inputs are less likely to be classified in the intended adversarial class, with their probability of success dropping steeply. Benign inputs are also affected but to a much smaller degree. 

We propose using the errors introduced by controlled undervolting of a chip to introduce sufficient perturbations to the model to disrupt adversarial inputs. The proposed system is deployed in three main phases, shown in Figure \ref{fig:globalflow}, as follows:  

\begin{figure*}[ht]
\minipage{\textwidth}
\hspace{2cm}
   \includegraphics[width=0.8\textwidth]{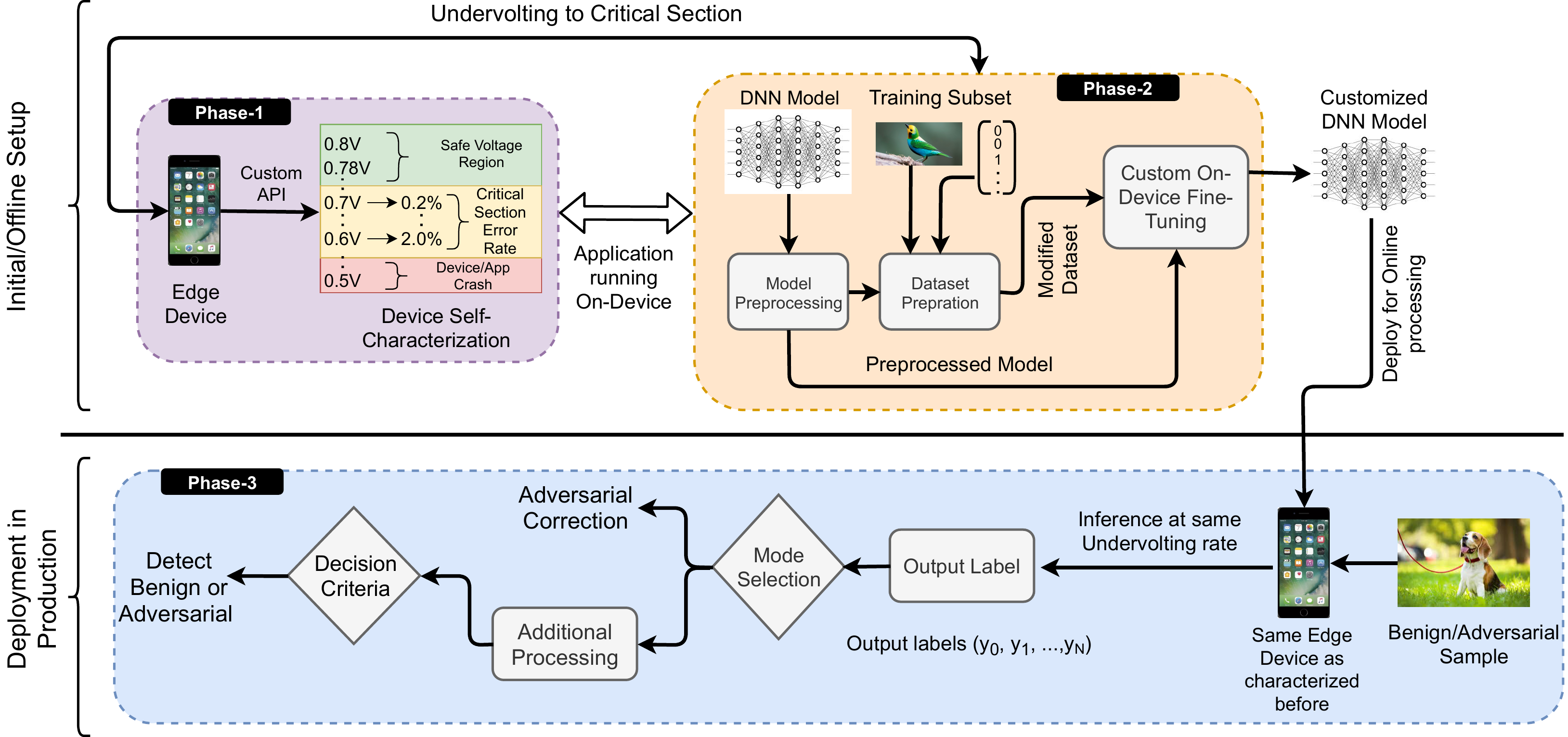}
\endminipage
\caption{\small Overview of proposed defense. A setup phase characterizes the voltage/error profile for each device (Phase 1), and performs on-device model fine-tuning (Phase 2). In production, the device is undervolted and a correction or detection algorithm is applied (Phase 3). }\label{fig:globalflow}
\end{figure*}

\subsection{Phase I: Offline Device Characterization}

In general, the voltage spectrum of a device can be classified into three regions: \textit{Safe, Critical and Crash}. In the \textit{Safe} region, the device operates normally, with no errors. In the \textit{Critical} voltage region the chip will experience occasional errors. Depending on the system and/or application it may continue to operate in the presence of these errors. Finally, the \textit{Crash} voltage is too low for the system to operate. Our defense utilizes voltages in the \textit{Critical} range.

\begin{figure}[htbp]
\centering
\vspace{20pt}%
     \includegraphics[width=\linewidth]{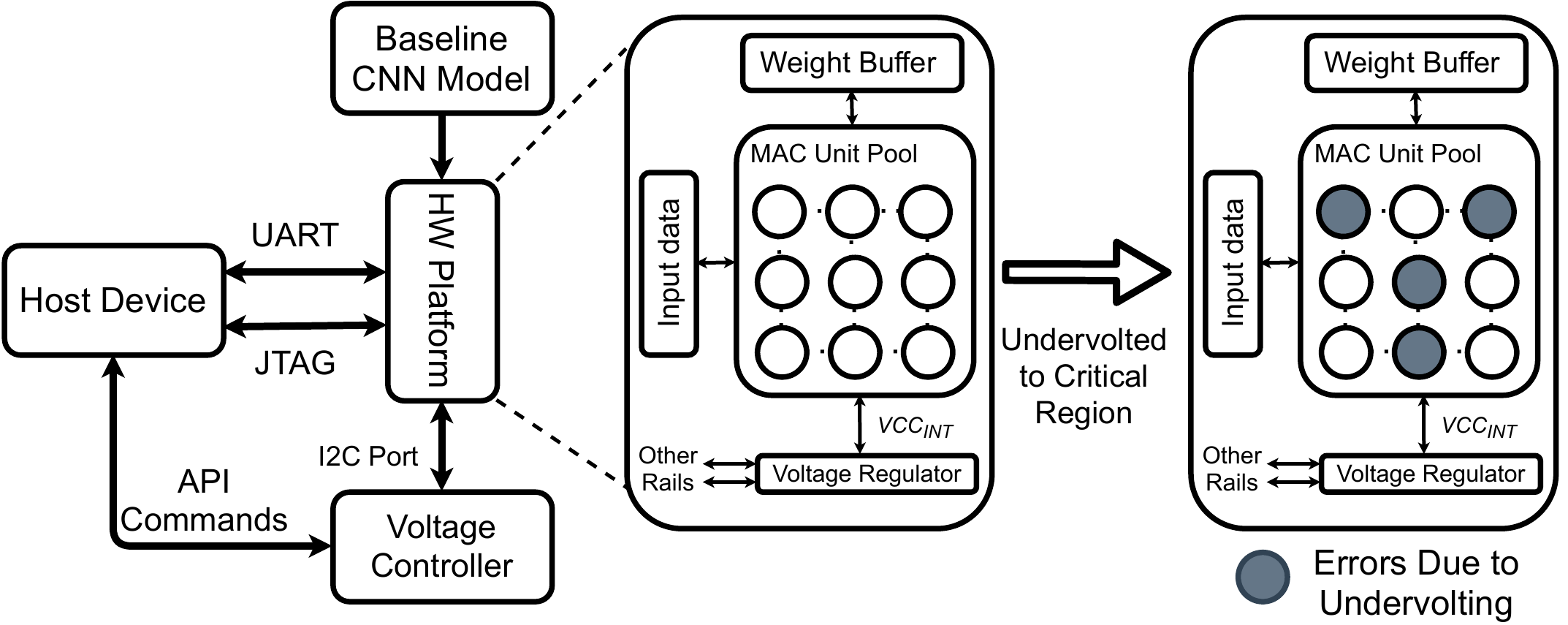}
  \caption{\small Characterizing undervolting effects on a hardware DNN accelerator. Error rates are recorded for multiple voltages within the critical region.}\label{fig:hwflow}
\end{figure}

Manufacturing process variation makes the response of each device to undervolting unique. As a result, the critical region resides at different voltage ranges for each chip. The target device therefore needs to be programmed to characterize these regions after manufacturing. This characterization can be performed by the device manufacturer, and it involves progressively lowering the supply voltage and running a built-in self test application designed to detect compute or memory errors (Figure \ref{fig:hwflow}). Using the error rate profile, the device can undergo controlled undervolting in the critical region during execution.

\subsection{Phase II: At Setup on-Device Tuning}





Our analysis shows that the accuracy of the classification of benign inputs can also be affected by undervolting. To reduce this impact we develop a lightweight on-device fine-tuning process to refine the model to account for errors. This solution takes advantage of the fact that undervolting errors occur in the same functional units of a given chip. In accelerator designs that use mostly static mapping of computation to hardware, this means that errors affect mostly the same model nodes across runs. Fine-tuning the \textit{pre-trained} classifier model in the presence of errors allows the model to adapt and recover most of the accuracy loss. An important constraint on the fine-tuning process is to keep its overhead very low, given that it needs to run on edge devices. This means the fine-tuning should rely on a small training set, to ensure low storage and communication overhead, as well as a small number of training iterations. 


Given these constraints, a small subset of less than 0.6\% set is sampled randomly from the full training dataset of the model. We explored a range of parameters for the on-device fine tuning, including the number of training epochs and training set size. We choose parameter values at which fine tuning begins to show diminishing returns. 

\subsubsection{Model Preprocessing} Modern CNNs consist of a Softmax layer which is generally used as the last activation function of a neural network to normalize the output vector $Z$ produced by the last hidden layer of a CNN, called \textit{logits}, to a probability distribution over predicted output classes. A single unit of a standard softmax function is denoted as:

\begin{displaymath}
\sigma_{i}= \frac{e^{z_{i}/T}}{\sum\limits_{j=1..N}e^{z_{j}/T}}
\end{displaymath}

where $Z = (z_{1}, z_{2},..., z_{N})$, are the $N$ logit outputs corresponding to N classes in the dataset and $T$ denotes the \textit{temperature} parameter. In general, the $T$ value is set to 1 producing a discrete distribution between 0 and 1. We set $T$ to a higher value in order to increase the uncertainly in the probability distribution vector leading the vector components to converge towards $1/N$ \cite{papernot2016distillation}. The softmax layer in the \textit{pre-trained} model is modified with a custom softmax-based activation function. The value of T is model-specific and can be refined along with other parameters such as learning rate and number of epochs, to achieve fast and accurate fine-tuning without overfitting. 


\subsubsection{Dataset Processing} 
\begin{figure}[ht]
\centering
     \includegraphics[width=\linewidth]{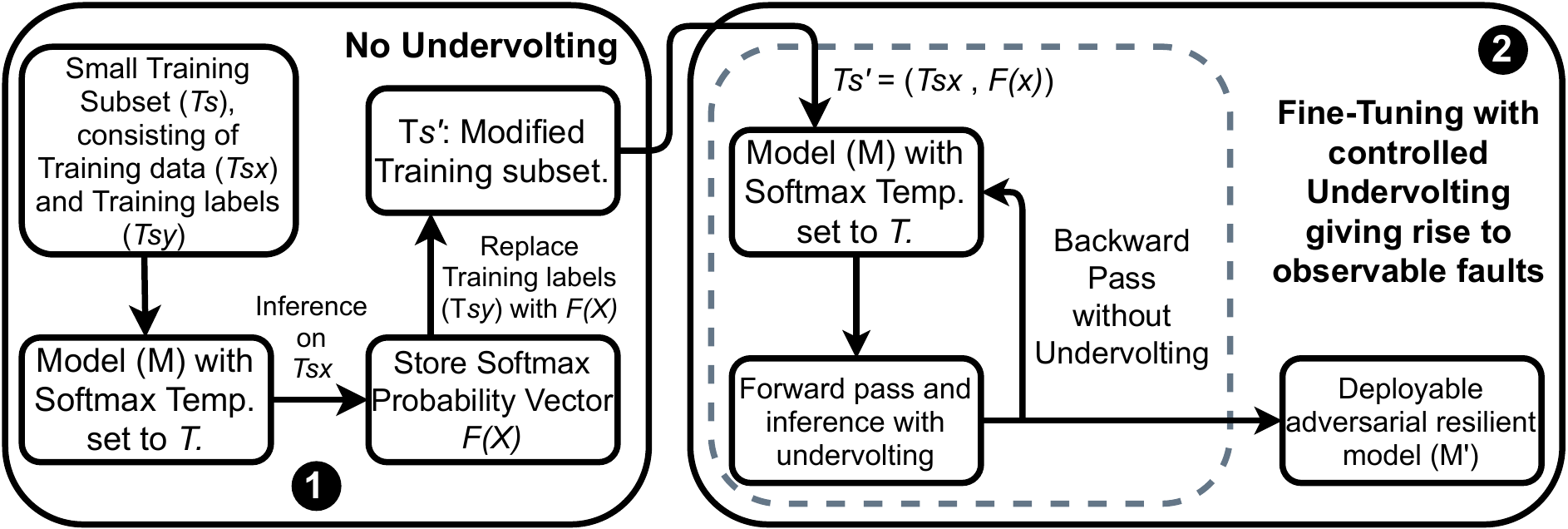}
  \caption{\small Training subset processing (1) and undervolting-aware model fine-tuning (2).}\label{fig:datasetfinetune}
\end{figure}

A small training dataset is required for fine-tuning. The dataset selection and processing is highlighted in Step \circled{1} of Figure \ref{fig:datasetfinetune}. A training set consists of $(X, y)$ tuple where
$X=(x_0, x_{1},.., x_{K})$ denoting $K$ input samples/images and $y=(y_{1i}, y_{2i},.., y_{Ki}  \vert i = 1, 2,.., N)$, denoting $N$ class labels for each of the $K$ inputs. The values for these $N$ class labels are binary, with only the correct class having a value of 1, with the rest 0. The idea of this step is to use the class probability vectors (\textit{soft labels}) produced by the softmax layer instead of the discrete-binary or \textit{hard} class labels to fine-tune the model. Using the probability vectors as the \textit{Y} values is beneficial as they contain additional information about each class instead of simply providing the correct class label. The intuition is that when a model is being trained, the knowledge is encoded in both the weight parameters and the probability vectors. 

The second fine-tuning step is \circled{2} in Figure \ref{fig:datasetfinetune}. The targeted chip is undervolted in order to surface the hardware errors and their effect on the model under training. Given that the error distribution does not change significantly over time, the idea is to adjust the model to account for this distribution. The goal is to recover some of the classification accuracy loss caused by the errors to benign inputs. The pre-trained model is fine-tuned with the modified dataset under errors. For a model M with features $\theta$ and a given set of sample $(x, P(x) \vert x\in X)$, where $x$ is the input sample, and $P(x)$ are the soft labels, the goal of this phase is to solve the following optimization problem: 

\vspace{-8pt}
\begin{displaymath}
\arg \min\limits_{\theta} - \frac{1}{|X|} \sum\nolimits_{x \in X} \sum\nolimits_{i \in N}P_{i}(x)\log M_{i}(x)
\end{displaymath}
This means for each sample $(x, P(x))$ we consider the log-likelihood $L(M, x, P(x))$ of model $M$ on $(x, P(x))$ and average it for the training set $X$. 

\begin{figure}[ht]
\centering
     \includegraphics[width=\linewidth]{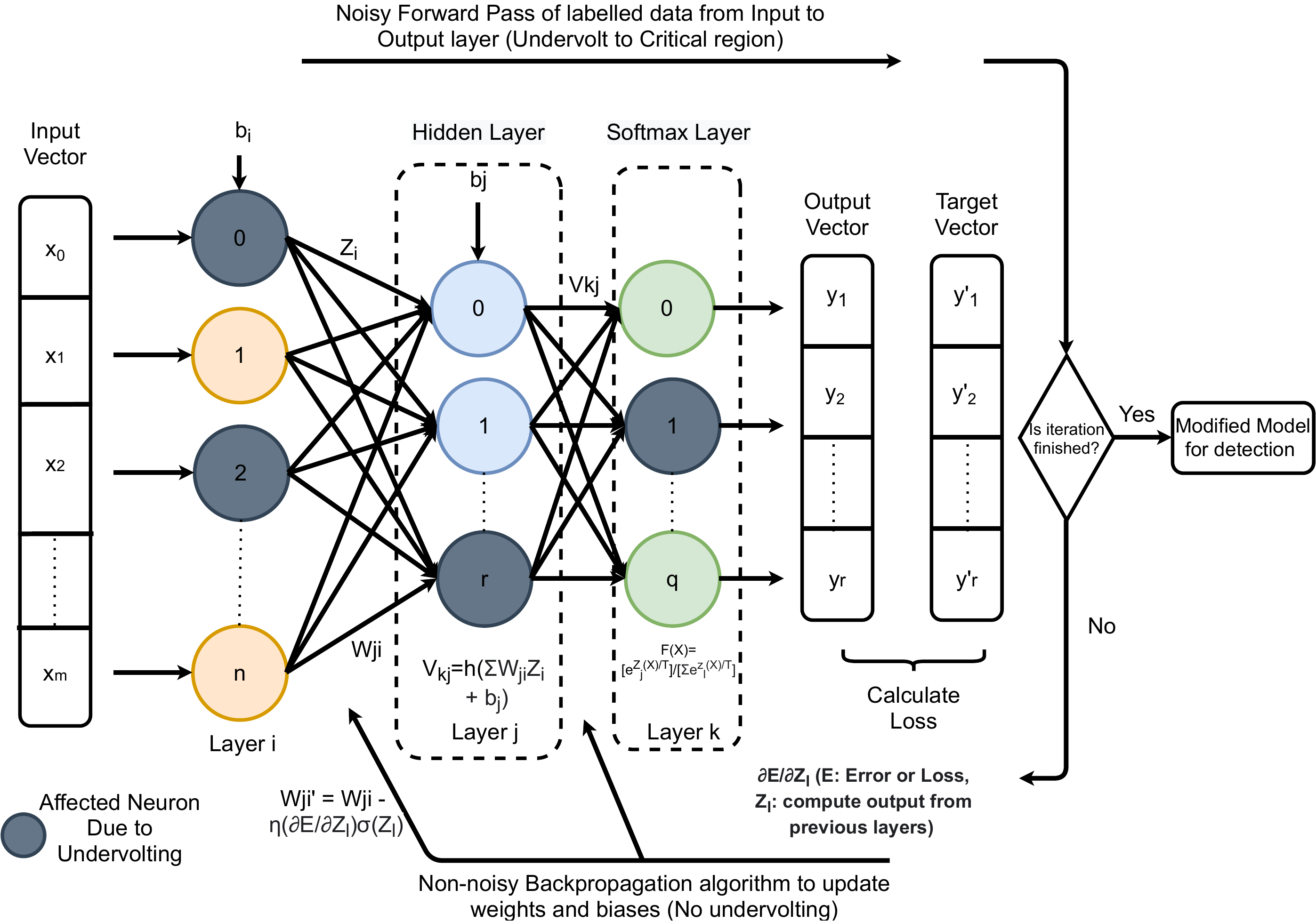}
  \caption{\small On-device model fine-tuning flow in the presence of undervolting-induced errors.}\label{fig:finetune}
\end{figure}

The fine-tuning process requires two passes. The forward pass predicts the output class labels based on some input data. In the backward pass the predicted output is compared with the actual target output. The loss is calculated for that iteration, and is used to update the weights and biases accordingly. The process repeats until it reaches the iterations limit or when the model achieves a threshold accuracy. In our design only the forward pass is undervolted, while the backward pass is performed at nominal voltage to ensure error-free backpropagation of updated parameters. Figure \ref{fig:finetune} illustrates this process. Once the fine-tuning is complete, the updated model is ready to be deployed in production.

\subsection{Phase III: Runtime Adversarial Correction/Detection}

The final phase of our proposed design consists of the deployment of the modified model in the adversarial defense mechanism. We examine two designs, with different effectiveness and overhead tradeoffs: (1) adversarial correction, which accurately classifies a subset of the adversarial inputs with no performance overhead and lower energy then the baseline, and (2) adversarial detection, which detects $>$90\% of adversarial inputs at the cost of higher performance overhead. We describe the two approaches as follows:


\subsubsection{Adversarial Correction} 
\begin{figure}[htbp]
\centering
     \includegraphics[width=1.0\linewidth]{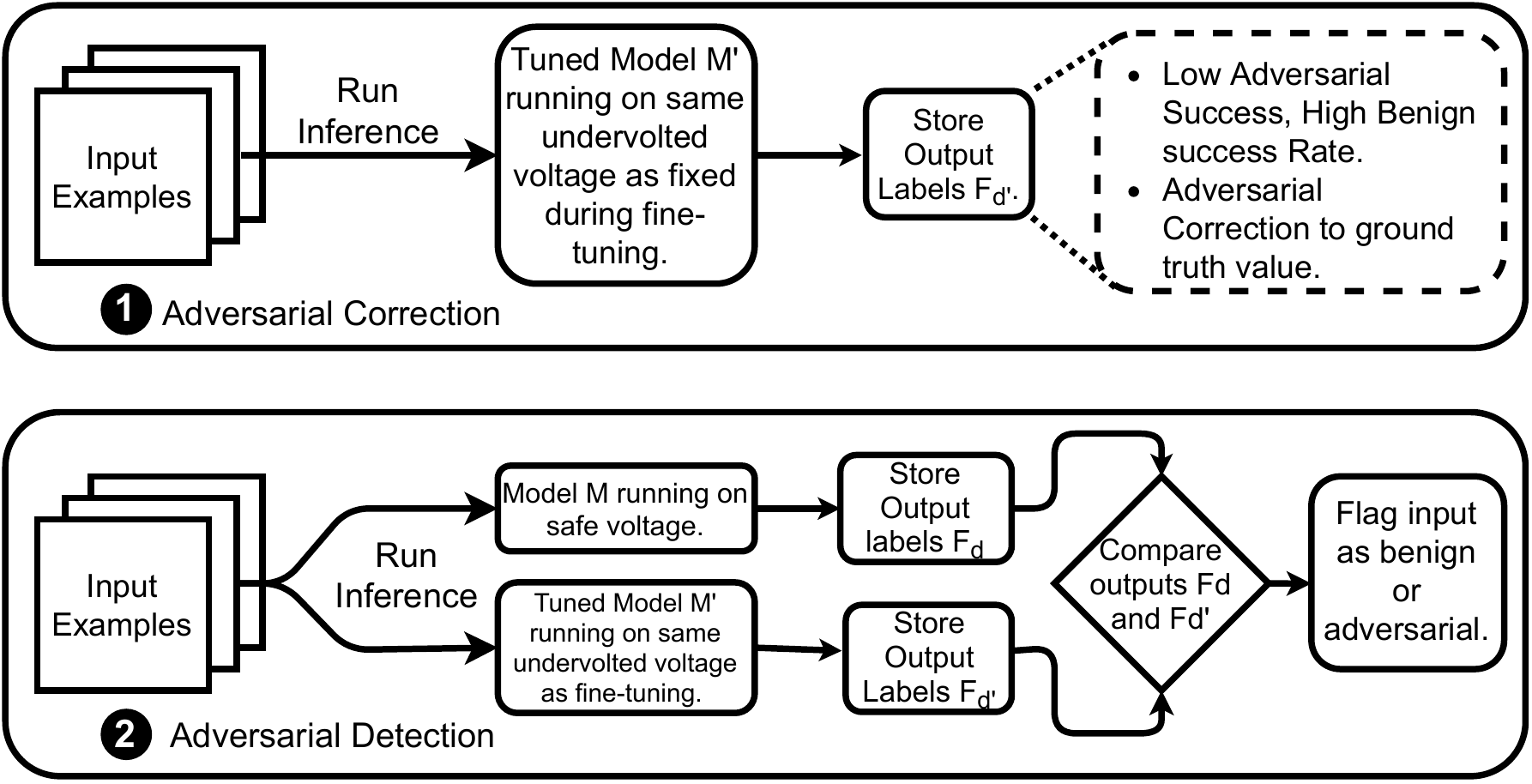}
  \caption{\small Runtime adversarial correction and detection mechanisms.}\label{fig:CorrDetect}
\end{figure}

The goal of a model or a classifier m, having logit values as $k_{y}$, can be given as $F(x)=arg max_{y}k_{y}(x)$, where $x$ is an input to the model m. Now, for an adversarial that has been perturbed from the original input $x$ as ($x+\Delta x$), the goal for an adversarial correction can be summarized as, $F(x+\Delta x)=y=F(x)$, where $y \in (1,...,K)$, denoting the class labels. By simply running the fine-tuned model on an undervolted chip, the associated errors will induce the classifier to correctly classify a majority of the adversarial inputs. Figure \ref{fig:CorrDetect} \circled{1} illustrates the adversarial correction flow. This approach has no performance overhead and, because of undervolting, has lower energy than the unprotected baseline. The correction rate is around 50-60\% of the adversarial inputs we test. 

\subsubsection{Adversarial Detection} In order to detect if an input is benign or adversarial, our approach compares the classification output of a reference inference pass at nominal voltage with an undervolted inference pass for the same input. In the first pass, a model with the same characteristics and hyperparameters as the baseline model is selected to get the classification result for the input example \textit{without} undervolting. In the second pass, the modified model is used to get the classification result, \textit{with} undervolting. Both these inference results are compared and, if the results match, the input image is labeled as legitimate.  Otherwise, it is flagged as adversarial. Figure \ref{fig:CorrDetect} \circled{2} illustrates the adversarial detection process. The intuition behind this approach is that undervolting is more likely to change the classification (relative to reference run) of the adversarial inputs. Benign inputs are less likely to be affected and their classification will match the reference. We show that this approach is 80-90\% effective at detecting adversarial inputs. However, since it requires an error-free reference run, this approach doubles the inference latency -- although the energy overhead is only about 70\% due to the undervolting power savings.

\section{Methodology}
\label{sec:method} 

\subsection{Evaluation platform}

\begin{figure}[htpb]
\centering
     \includegraphics[width=0.95\linewidth]{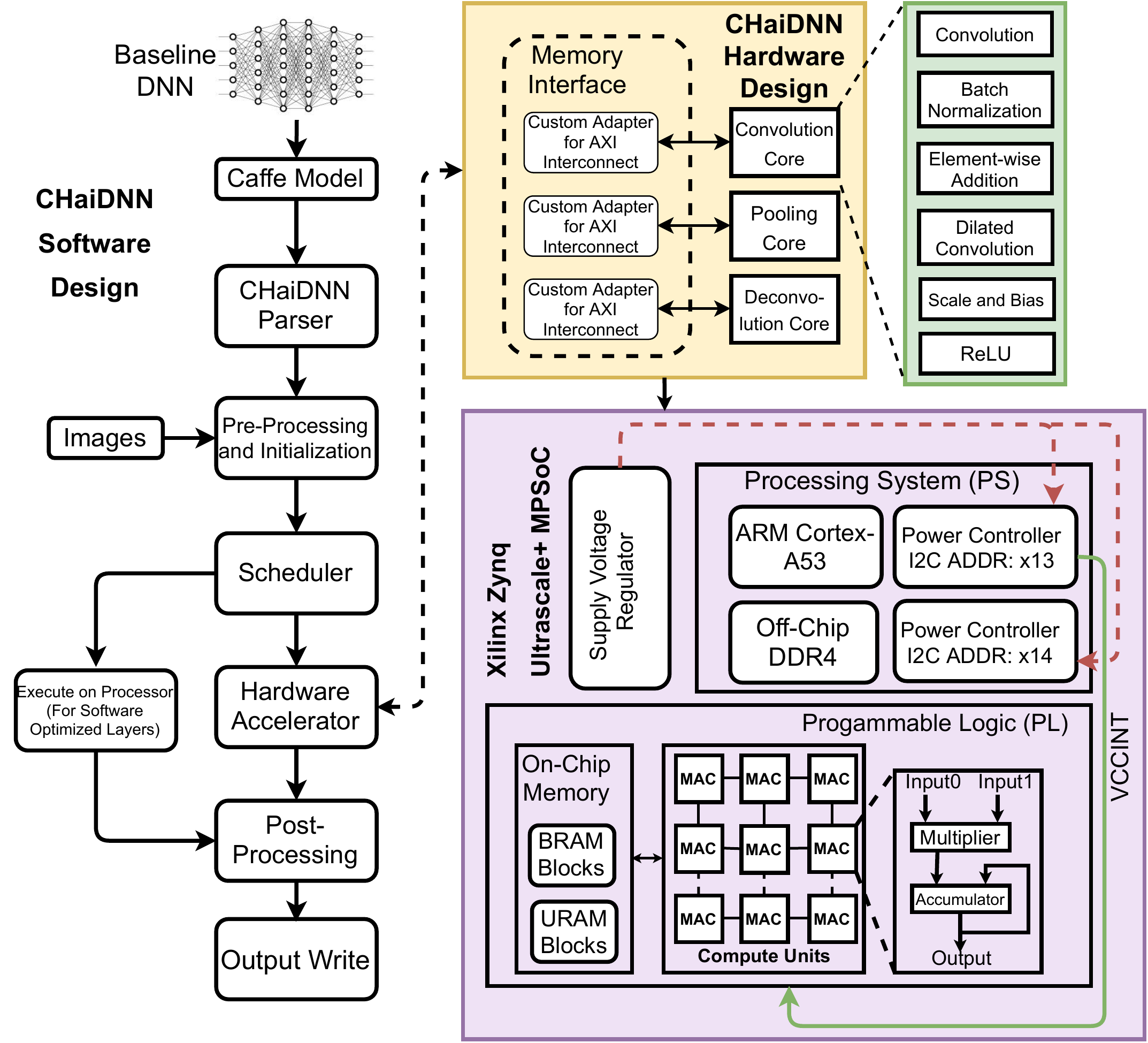}
  \caption{\small CHaiDNN accelerator design on ZCU104 platform.}\label{fig:chaidnn_base}
\end{figure}

We use CHaiDNN\cite{chaidnn}, a Deep Neural Network acceleration library designed for Xilinx UltraScale MPSoCs as our evaluation platform. CHaiDNN integrates both system software and a hardware accelerator (Figure \ref{fig:chaidnn_base}), supports most of the widely used CNN models, and custom networks and layers.

\subsection{FPGA Platform} 

We deploy CHaiDNN on the Xilinx Zynq Ultrascale+ ZCU104 FPGA platform \cite{ZCU104}, which uses the XCZU7EV-2FFVC1156 MPSoC. The device includes a quad-core Arm Cortex-A53 applications processor (APU) and a dual-core Cortex-R5 real-time processor (RPU), and a graphics processing unit. The MPSoC is fabricated in a 16nm FinFET+ process technology. The Programming Logic (PL) features about 504K logic blocks (LUTs), 11Mb of Block RAM (BRAM), 27Mb of UltraRAM (URAM) and 1728 signal processing (DSP) slices. The board is also equipped with a 2GB 64-bit DDR4 off-chip memory. 


\subsection{Datasets and Classifier Models}

\begin{figure}[h]
\centering
     \includegraphics[width=\linewidth]{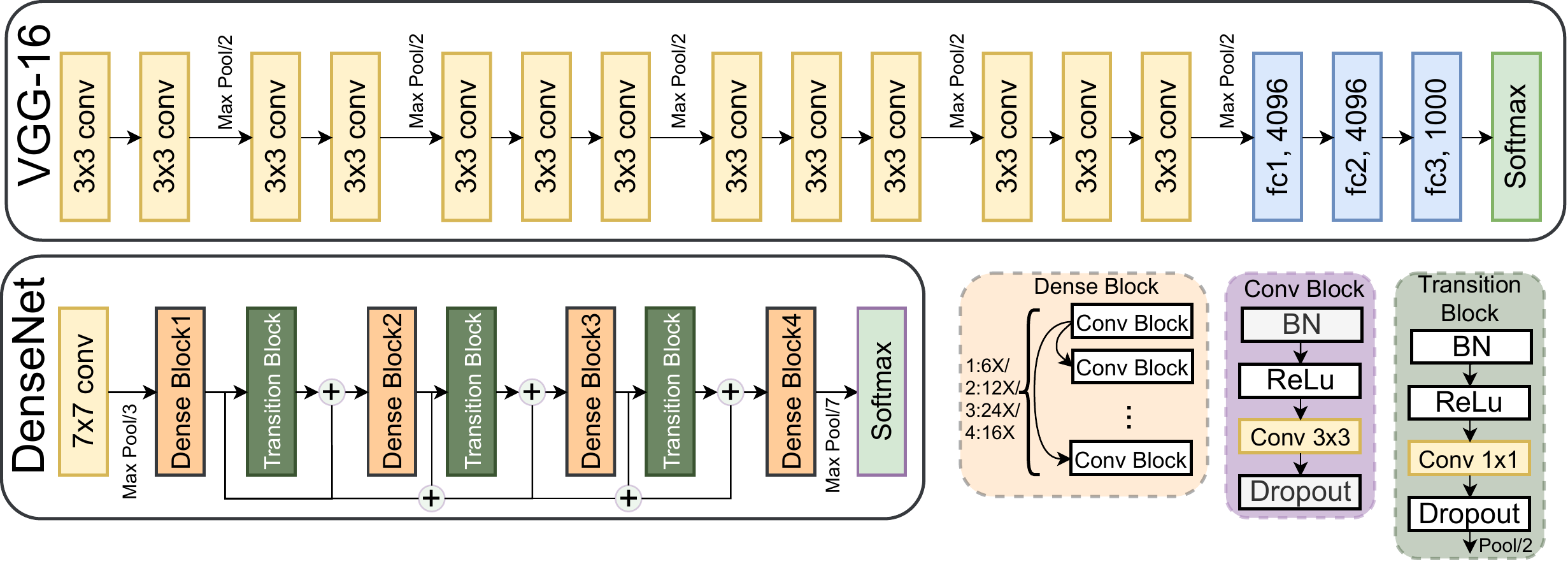}
  \caption{\small VGG16 and DenseNet classifier architectures.}\label{fig:models}
\end{figure}

We evaluate our defense on two popular convolution-based deep neural networks datasets and models: 

1. \textbf{CIFAR-10 on DenseNet}: The CIFAR-10\cite{cifar10} dataset consists of 60,000 32$\times$32 color images in 10 classes, with 6K images per class. There are 50K training images and 10K test images. We use a DenseNet\cite{huang2018densely},\cite{densenet} model to evaluate the CIFAR-10 dataset. As Figure \ref{fig:models} shows, DenseNet is composed of an initial convolution layer followed by Dense and Transition blocks and a softmax classifier. In the Dense Blocks, the layers are densely connected. 

2. \textbf{ImageNet on VGG16}: The ImageNet\cite{krizhevsky2012imagenet} project is a large visual database designed for use in visual object recognition software research. The most used subset of ImageNet is the ImageNet Large Scale Visual Recognition Challenge (ILSVRC)\cite{ILSVRC15} image classification and localization dataset. This dataset spans 1K object classes and contains 1,281K training images, 50K validation images and 100K test images. We use the VGG16\cite{simonyan2015deep} model to classify the ImageNet dataset. As Figure \ref{fig:models} shows, VGG16 consists of 16 convolution layers interleaved with activation layers, followed by 3 fully connected layers and a softmax layer for classification. 


\subsection{Attack Types and Datasets}

We evaluate our defense on the attacks summarized in Table \ref{tab:attack_stats}. We use two untargeted attacks: FGSM, Deepfool and four targeted attacks: $CW_{L0}$, $EAD_{L1}$, $CW_{L2}$, $CW_{L\inf}$. Targeted attacks aim to misclassify an input into a target class. We deploy two different targets: the \textit{least-likely(LL)} class (\textit{$t=min(\hat{y})$}) and the \textit{Next} class ($t=L+1~mod~\#classes$), where $t$ is the target class, $L$ is the sorted index of top ground truth classes, and $\hat{y}$ is the prediction vector of an input image. We select a total of 1K legitimate samples and 100 adversarial samples for each attack method from each of the two dataset.


\begin{table}
\caption{\small Evaluated adversarial attack characteristics.}
\label{tab:attack_stats}
\small
\centering
\scalebox{0.7}{
\begin{tabular}{|c|c|c|c|c|c|c|c|c|}
\hline

{\multirow{2}{*}{Dataset}}&{\multirow{2}{*}{Attack}} & \multirow{2}{*}{{Mode}} & {\multirow{2}{*}{Success rate}} & {\multirow{2}{*}{ Prediction Conf.}}& 
\multicolumn{3}{c|}{ Distortion}  \\

\cline{6-8}&{}&
{}&{}&{}&{\large $L_0$}  & {\large $L_2$} & {\large $L_{\infty}$}\\
 
\hline

&{\textbf{$FGSM$}} &    - & 99\%&63.9\%&94\%&3.1&0.008\\
\cline{2-8}

\multirow{4}{*}{\rotatebox{90}{\scalebox{1.6}{$ImageNet$}}}&\multirow{2}{*}{\textbf{$CW_{L0}$} } &LL& 100\%&84.4\%&42.4\%&13.65&0.96\\    && Next & 97\%&92.9\%&42.2\%&10.44&0.94\\

\cline{2-8}

&\multirow{2}{*}{\textbf{$CW_{L2}$} } &LL& 97\%&75.25\%&54.3\%&1.027&0.031\\
&&Next & 90\%&76.3\%&32.2\%&0.66&0.019\\

\cline{2-8}

&\multirow{2}{*}{\textbf{$CW_{L\infty}$}} &LL& 100\%&91.8\%&99.9\%&3.05&0.01\\

&&Next & 100\%&94.7\%&100\%&2.27&0.01\\
\cline{2-8}

&\multirow{2}{*}{\textbf{$EAD_{L1}$}}
&LL& 100\%&100\%&54.7\%&3.56&0.29\\
&&Next& 100\%&95.5\%&43.8\%&3.87&0.7\\
\cline{2-8}

&{\textbf{$DeepFool$}} &    - & 100\%&79.59\%&98.4\%&0.726&0.027\\

\cline{1-8}
\\
\hline

&{\textbf{$FGSM$}} &    - & 100\%&84.85\%&99.7\%&0.863&0.016\\
\cline{2-8}

\multirow{4}{*}{\rotatebox{90}{\scalebox{1.2}{$CIFAR-10$}}}&\multirow{2}{*}{\textbf{$CW_{L0}$} } &    LL& 100\%&97.60\%&2.4\%&2.530&0.712\\
&& Next & 100\%&98.19\%&1.9\%&2.103&0.650\\
\cline{2-8}

&\multirow{2}{*}{\textbf{$CW_{L2}$} } &    
LL& 100\%&97.35\%&85.5\%&0.358&0.042\\
&&Next & 100\%&97.90\%&76.8\%&0.288&0.034\\

\cline{2-8}

&\multirow{2}{*}{\textbf{$CW_{L\infty}$}} &    
LL& 100\%&97.79\%&99.5\%&0.527&0.014\\
&&Next & 100\%&98.22\%&99.0\%&0.446&0.012\\
\cline{2-8}

&{\textbf{$DeepFool$}} &    - & 98\%&73.45\%&99.5\%&0.235&0.028\\
\hline

\end{tabular}
}
\end{table}

\subsection{Undervolting Methodology}

\begin{figure}

\minipage{0.5\linewidth}%
  \includegraphics[width=0.9\linewidth]{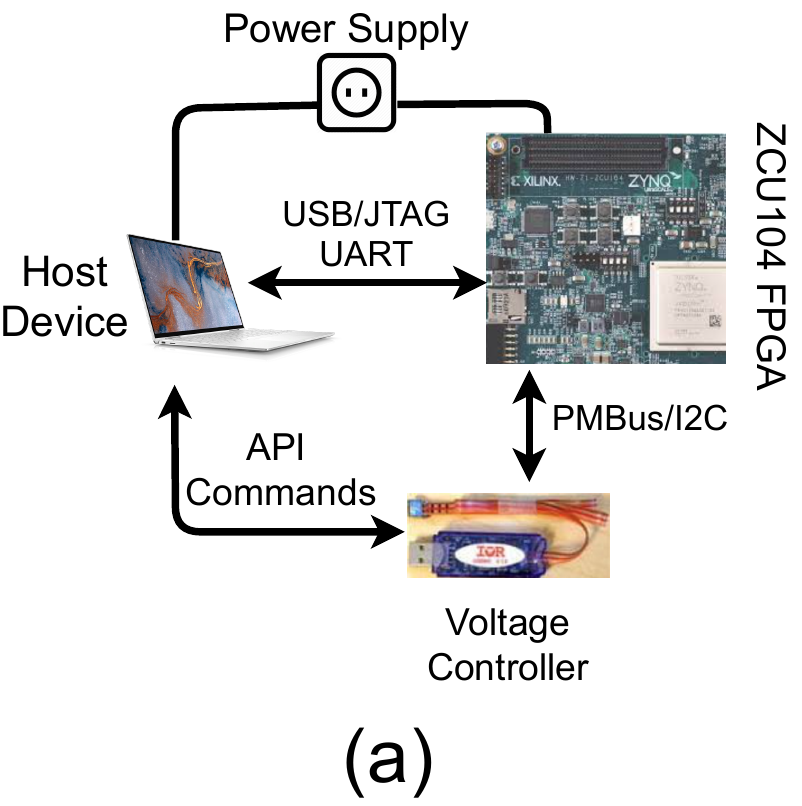}
\endminipage
\minipage{0.5\linewidth}  
  \includegraphics[width=0.9\linewidth]{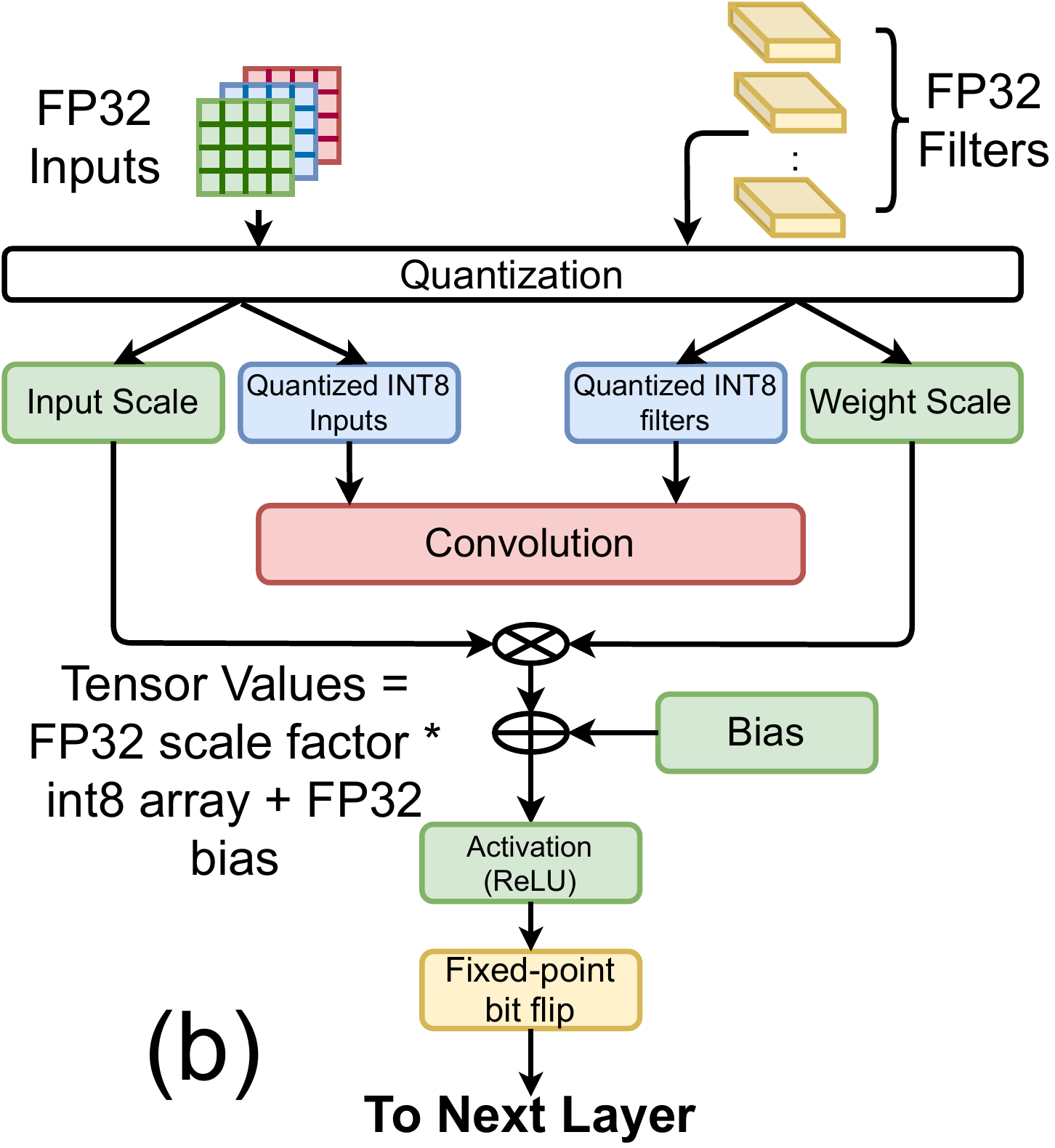}
\endminipage
\caption{\small Hardware platform setup (a) and software model flow (b).}\label{fig:fpgasoft}
\end{figure}

We perform undervolting characterization on the FPGA device using an external voltage controller, the Infineon USB005 \cite{infineon}, which is connected to the board using an I2C cable. The provided PowerIRCenter GUI enables reading and writing the different voltage rails to the board. We can also monitor the power consumption and the temperature of the chip using this GUI. For this study, we focus on $VCC_{INT}$ voltage rail accessible using PMBus address 0x13. 
This voltage rail drives most of the programming logic components, including the DSPs and LUTs. The layout of our hardware platform setup is illustrated in Figure \ref{fig:fpgasoft}(a).


We gradually sweep the entire voltage range available to our device, while under load, until we reach the crash region. While undervolting in the critical region ($>=0.660V$ to $< 0.648V$) the CHaiDNN Scheduler application running on the CPU experiences frequent crashes. We attribute this behavior to be the shared voltage rail between the embedded CPU and the Programmable Logic units. In order to better control errors in the Programmable Logic units that contain the DNN accelerator, we inject random errors (as bit flips) directly in the Multiply-Accumulate (MAC) units of the ChaiDNN hardware. The MACs are the dominant components of the accelerator in terms of FPGA resource utilization. Our design deploys 256 such MAC units. 

\subsection{Software Simulation}


Since the CHaiDNN accelerator does not support training or fine-tuning, we evaluate the error injection and on-device fine-tuning on an Intel Core-i7 CPU@3.40GHz CPU using TensorFlow version 1.14\cite{tensorflow}, running the same models as CHaiDNN.  
We inject errors after every convolution layer of the models which includes activation functions. 
We also quantize the input and the weights to a fixed INT8 precision and then convolve them, followed by bias addition and activation. The reason for quantizing the inputs and weights is to match the CHaiDNN INT8 computations on the MAC units. 
Figure \ref{fig:fpgasoft}(b) illustrates the software error model and quantization process. 

\begin{figure}[htpb]
\centering
     \includegraphics[width=\linewidth]{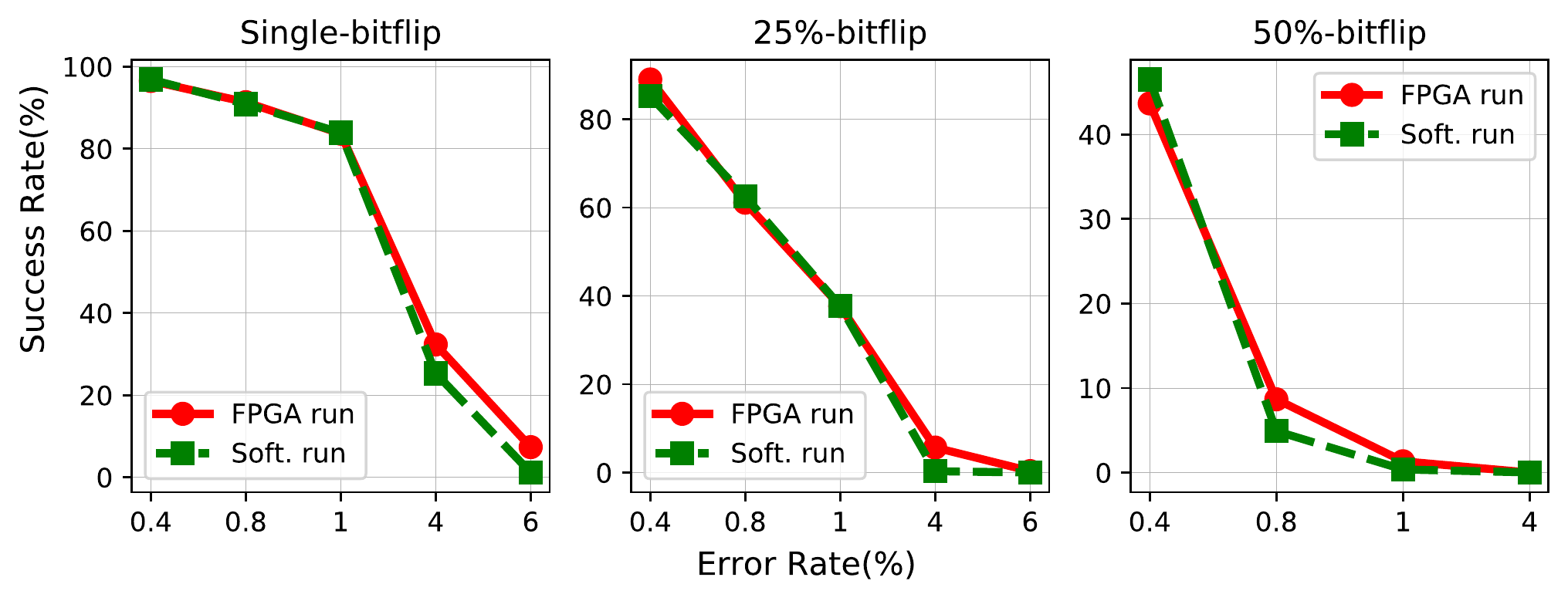}
  \caption{\small Undervolting effects on FPGA vs. software model.}\label{fig:FPGAvsSoft}
\end{figure}

This model is next fine-tuned following the process shown in Figure \ref{fig:finetune}. The error distribution is normal and random, with the final results averaged from multiple such distributions. To validate this simulation, we compare the accuracy loss due to undervolting effects on our FPGA platform running the accelerator and the software implementation on a sample benign test set.

Figure \ref{fig:FPGAvsSoft} shows the effects of undervolting the VGG16 model on the FPGA accelerator compared to the error injection in the software model. We can see that classification accuracy as a function of the error rate matches very well with the FPGA-based injection, across three different error models (single-bit error, 25\% bit flips and 50\% bit flips). This shows that, with some calibration, we can accurately replicate the effects of FPGA undervolting in the software model. This allows us to evaluate the custom fine-tuning as well as conduct additional sensitivity studies with respect to the error rates. 


\section{Evaluation}
\label{sec:eval}



\begin{figure*}[htbp]
\centering
\minipage{0.95\textwidth}
\begin{subfigure}[t]{\textwidth}
   \includegraphics[width=\textwidth]{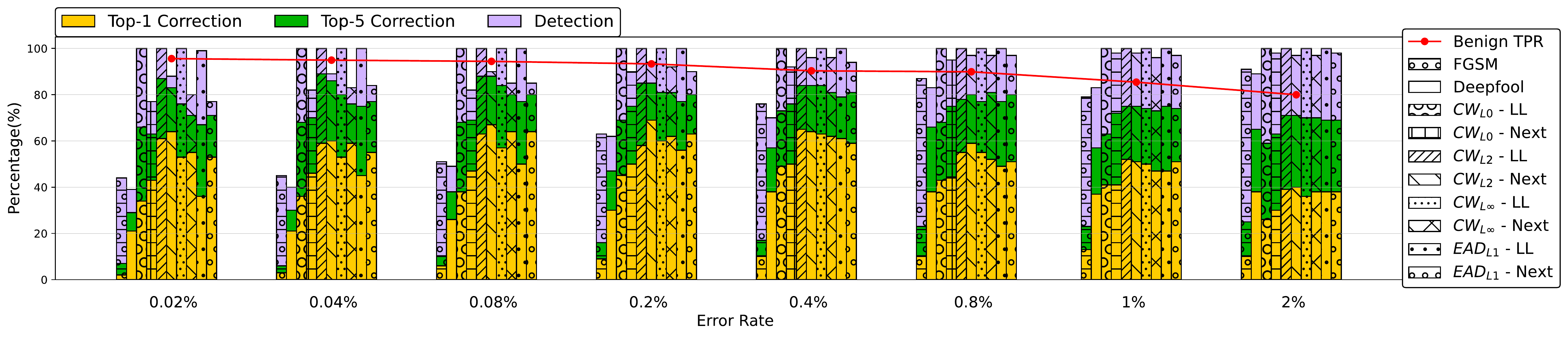}
\end{subfigure}
\endminipage\hfill
\minipage{0.95\textwidth}
  \begin{subfigure}[t]{\textwidth}
  \includegraphics[width=\textwidth]{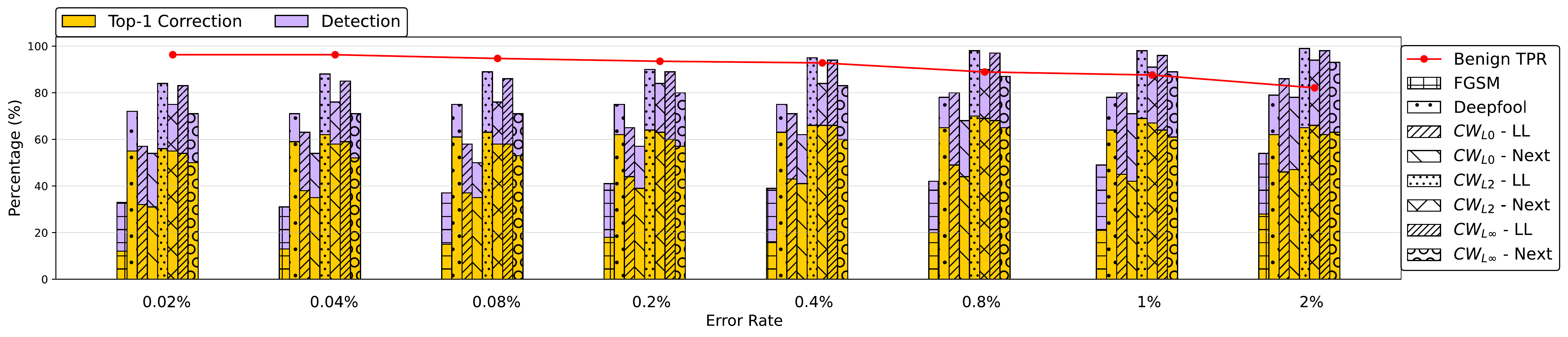}
\end{subfigure}
\endminipage
\caption{\small Adversarial detection/correction rate in ImageNet (top) and CIFAR-10 (bottom). True positive rate (TPR) for benign inputs.}\label{fig:overall_eval}
\end{figure*}

\subsection{Adversarial Detection and Correction}

Figure \ref{fig:overall_eval} shows the detection and the correction rates for all adversarial attacks under different simulated error rates for ImageNet and CIFAR-10 examples respectively. Our error model assumes, in this case, that 50\% of the bits in each affected output are flipped.


The error rate in the model will vary with the rate of undervolting, affecting the adversarial detection rate, which improves as the rate of error introduced in the model increases. For instance, with a 2\% error rate, the average detection rate is at 97\% compared to 81\% with a 0.02\% error rate in the Imagenet dataset. Adversarial correction is evaluated based on the predicted labels of the adversarial examples and comparing the Top-K of these labels with the ground truth of that adversarial. We show Top-1, and Top-5 correction rates for the ImageNet dataset and just Top-1 for CIFAR-10 since the latter only has 10 class labels compared to 1,000 class labels in ImageNet. Figure \ref{fig:overall_eval} indicates our defense mechanism is highly effective at reducing the success rate across multiple, diverse adversarial attacks. Our defense works best for low distortion, stronger attacks which are very difficult to detect, such as CW and EAD.

As expected, accuracy for the benign inputs decreases with the increase in error rate. However, the effect of the errors on benign inputs remains modest. The benign \textit{True Positive Rate (TPR)} reaches a low of 80\% for 2\% error rate. At the same time, it is above 90\% for lower error rates at which adversarial detection rates exceed 95\%. Fine-grain undervolting allows some control over the error rate. This enables the system to trade-off benign accuracy for adversarial detection rate, to adapt to the needs and constraints of the application.

\subsection{The Benefits of on-Device Fine-Tuning}

\begin{figure}
\centering
\minipage{0.6\textwidth}
\hspace{0.5cm}
    \includegraphics[width=0.65\textwidth]{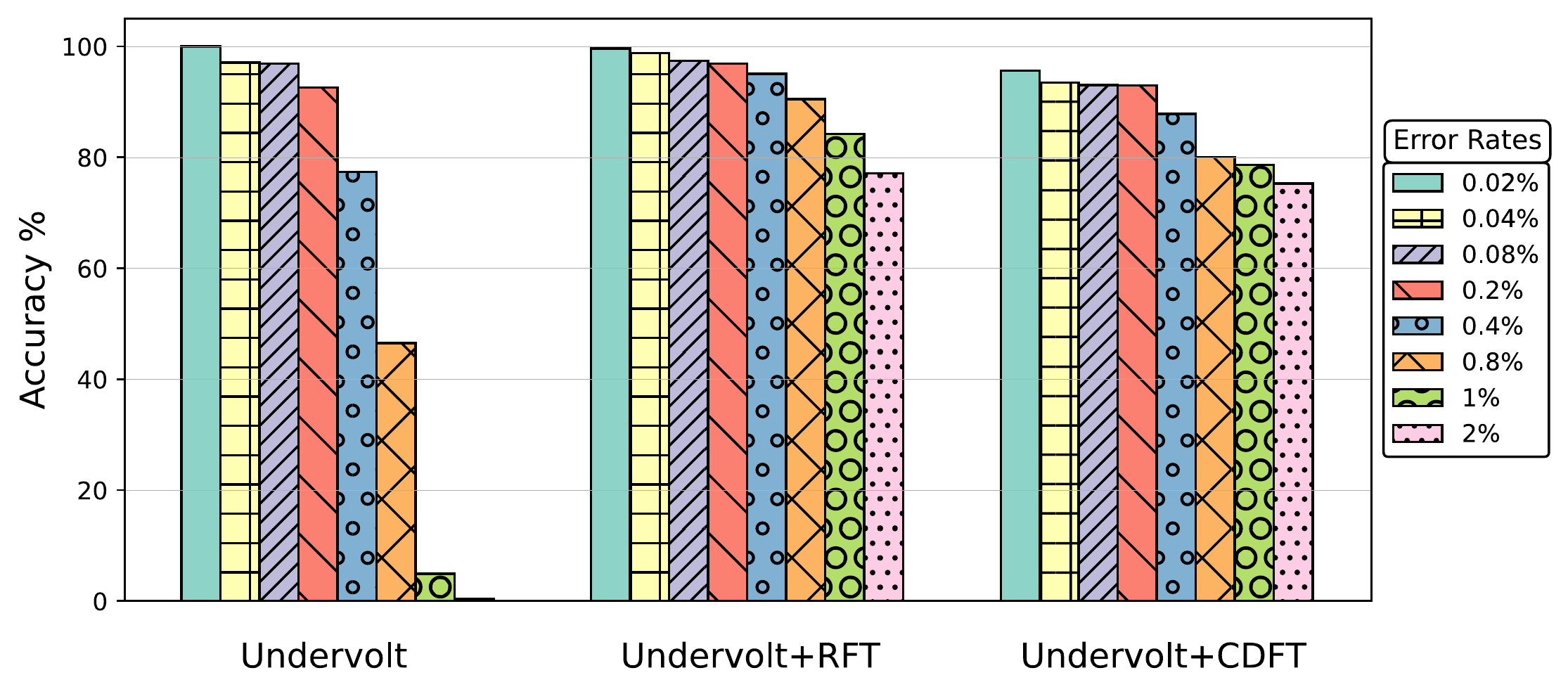}\par
\hspace{0.5cm}
    \includegraphics[width=0.65\textwidth]{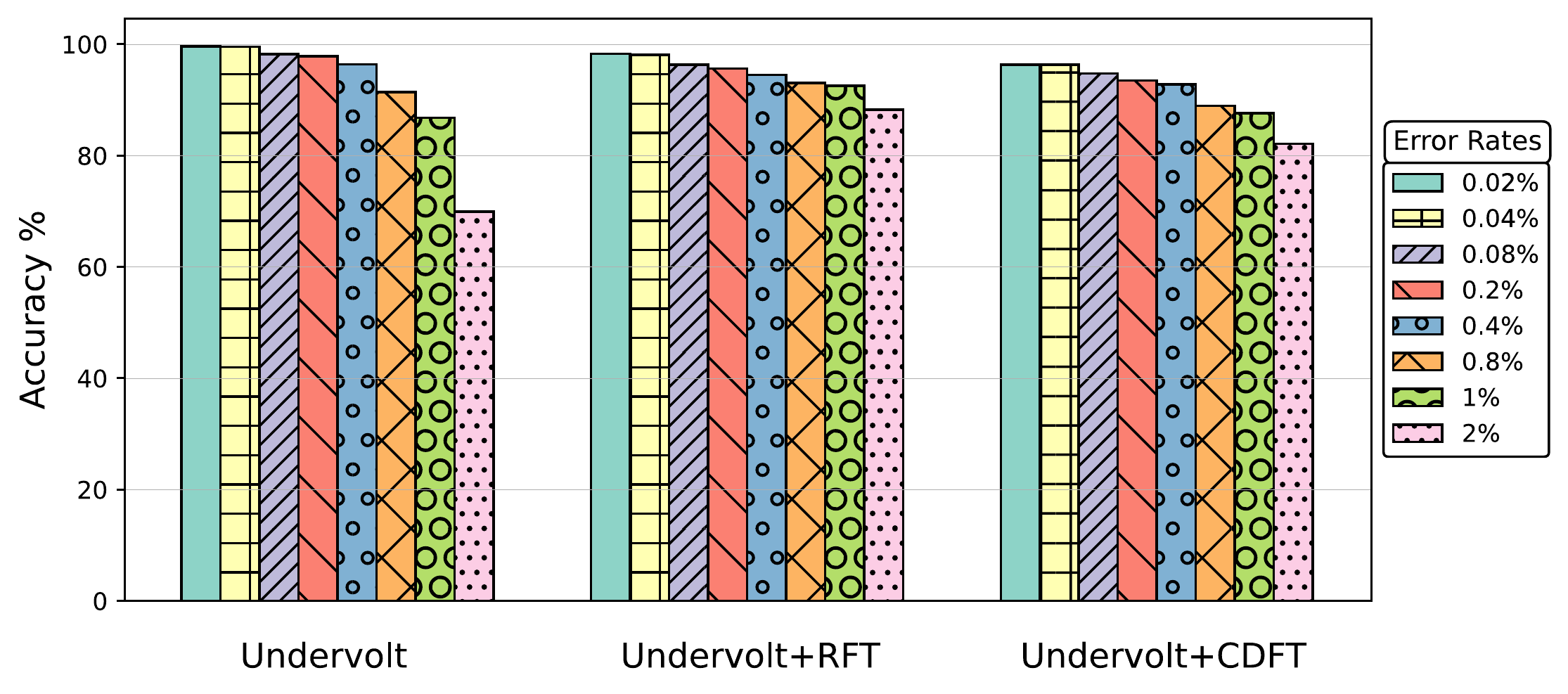}\par

\endminipage
\caption{\small Fine-tuning impact on benign inputs, ImageNet (top) and CIFAR-10 (bottom).}\label{fig:benign_imagenet_cifar}
\end{figure}

\begin{figure}
\minipage{0.6\textwidth}
\hspace{0.5cm}
    \includegraphics[width=0.65\textwidth]{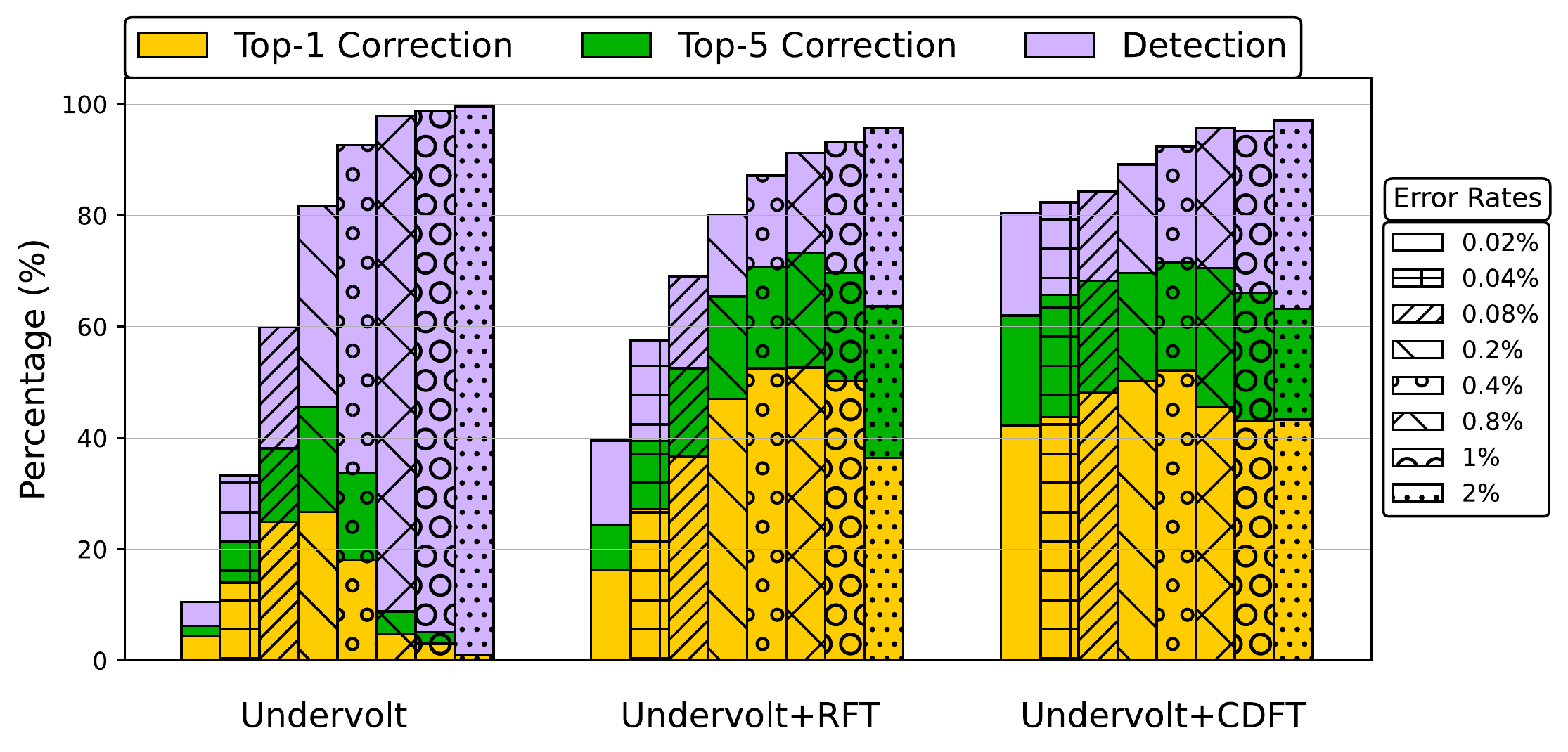}\par
\hspace{0.5cm}
    \includegraphics[width=0.65\textwidth]{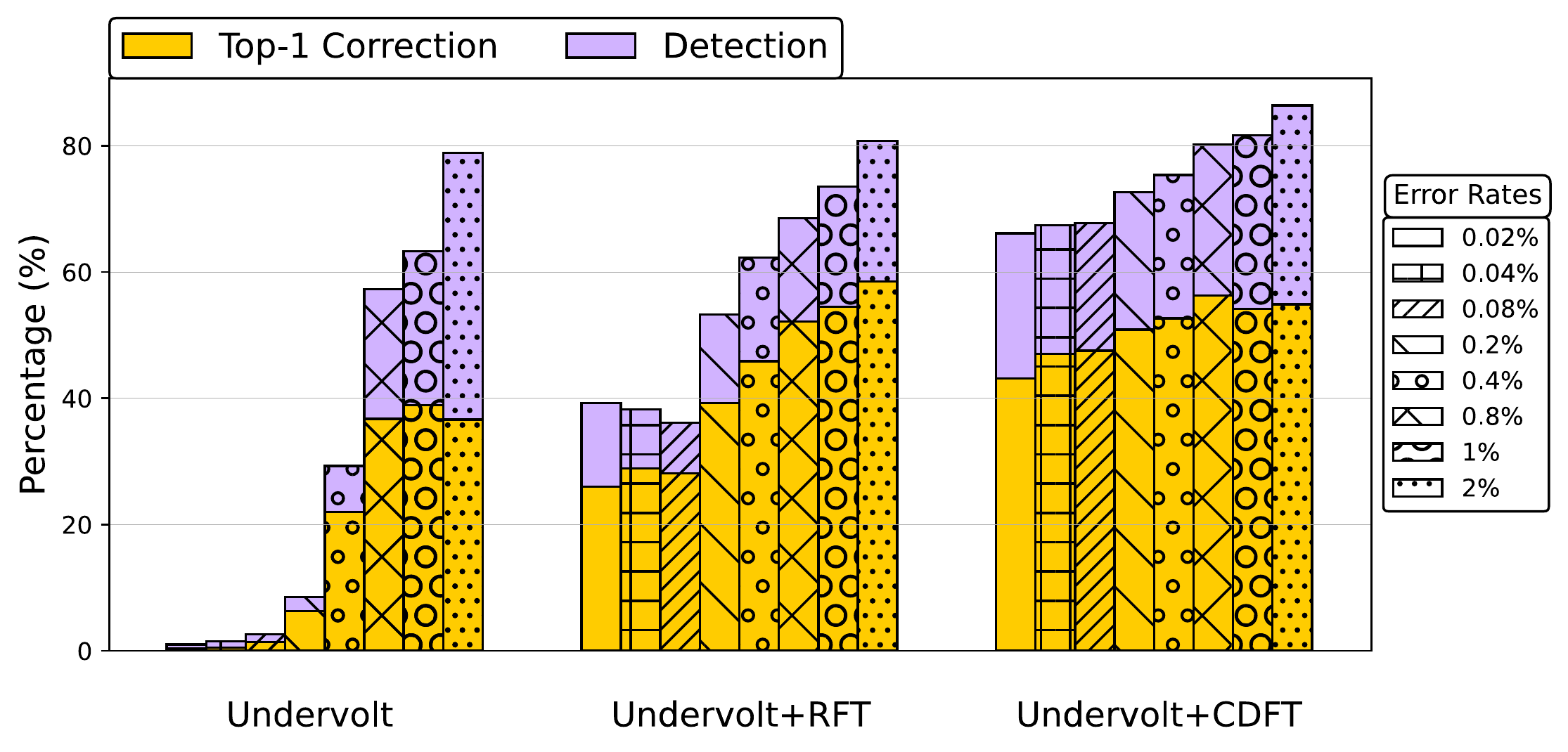}\par

\endminipage
\caption{\small Fine-tuning impact on adversarial inputs, ImageNet (top) and CIFAR-10 (bottom).}\label{fig:adv_imagenet_cifar}
\end{figure}




On-device fine-tuning is designed to mitigate some of the impact of undervolting errors on benign inputs and to improve detection/correction rate through custom distillation. Figure \ref{fig:benign_imagenet_cifar} shows the accuracy for benign inputs, and Figure \ref{fig:adv_imagenet_cifar} shows the detection and correction rates for adversarial inputs, averaged across all the attacks we study. We evaluate three designs at different error rates:


\subsubsection{Undervolt} is the core of our design. We evaluate the different models and datasets using just undervolting, with no fine tuning. We study how these errors affect the accuracy of the benign inputs and the success rate for different adversarial inputs. Figure \ref{fig:benign_imagenet_cifar} shows that, at low error rates, the accuracy of both models is quite high. However, at higher error rates the accuracy drops to almost 0\%. This shows that undervolting alone is only feasible at low error rates, where the adversarial detection rate is less than 80\%.


\subsubsection{Undervolt + Regular Fine-Tuning (RFT)} involves fine-tuning of the model with a small dataset. We observe that fine-tuning improves benign accuracy significantly at high error rates. For the ImageNet dataset, at the highest error rate of 2\%, the benign TPR increases from 0\% to 80\%. We see a 34.4\% average increase across all error rates. Fine tuning also improves adversarial correction rates, as Figure \ref{fig:adv_imagenet_cifar} shows. The average Top-1 correction rate increases 3.26$\times$ for ImageNet and 2.33$\times$ for CIFAR-10 vs.\ undervolting alone.



\textit{3) Undervolt + Custom Distillation-Based Fine-Tuning (CDFT):} adds a distillation to fine-tuning, as described in Section \ref{sec:design}. CDFT is particularly effective at lower error rates and complements undervolting well. For instance, in ImageNet with a 0.02\% error rate, we see a 2$\times$ increase in adversarial detection rate compared to fine-tuning alone. This is an important result since the adversarial detection/correction rate is significantly lower at low error rates. By adding distillation, we are able to raise the adversarial detection relative to fine-tuning alone, to an average of 90.1\% from 76.2\% for ImageNet and to 77.34\% from 56.2\% for CIFAR-10.

\begin{table*}
\caption{\small Detection and correction (Top-1) rate of multiple attacks and variants applied to samples from the ImageNet dataset.}
\label{tab:imagenet_attack_succ}
\small
\centering
\scalebox{0.79}{
\minipage{1.2\linewidth}
\begin{tabular}{|c|c|c||c|c||c|c||c|c|c||c|c|c|c|c|c|}
\hline

\multirow{4}{*}{Bit flips}&
\multirow{4}{*}{ Technique}&
\multirow{4}{*}{ Benign TPR(\%)} & \multicolumn{2}{c||}{$L_0$ attacks} & \multicolumn{2}{c||}{$L_1$ attacks} & \multicolumn{3}{c||}{$L_2$ attacks} & \multicolumn{3}{c|}{$L_{\infty}$ attacks}  \\

\cline{4-13}

&&&
\multicolumn{2}{c||}{CW}&
\multicolumn{2}{c||}{EAD} &
\multirow{2}{*}{Deepfool(\%)}&\multicolumn{2}{c||}{CW}&\multirow{2}{*}{FGSM(\%)}&\multicolumn{2}{c|}{CW} \\

\cline{4-7}
\cline{9-10}
\cline{12-13}
&&&LL(\%)&Next(\%)&
LL(\%)&\multicolumn{1}{c||}{Next(\%)}&
&LL(\%)&\multicolumn{1}{c||}{Next(\%)}&
&LL(\%)&Next(\%)\\

\hline

&Undervolt&25.3&100/5&100/7&100/5&100/8&98/8&100/6&100/9&87/1&100/4&100/9 \\


Single &Undervolt+RFT&89.3&100/51&92/51&100/53&88/61&64/37&100/67&96/63&67/8&100/62&96/60 \\

&Undervolt+CDFT&85.42&100/52&93/57&100/62&95/62&79/41&100/66&98/69&79/12&100/66&97/63\\

\hline
&Undervolt&62.5&100/12&89/22&98/15&88/26&78/18&100/27&93/30&1/5&100/30&89/19\\

25\%&Undervolt+RFT&91.2&100/39&82/51&98/45&85/59&59/26&100/57&93/59&48/5&100/64&88/55 \\

&Undervolt+CDFT&87.6&100/43&89/52&100/54&91/59&66/33&100/57&95/60&63/9&100/59&91/58\\
\hline

&Undervolt&77.4&100/13&91/19&100/16&93/17&85/17&100/23&93/28&59/4&100/19&97/25\\

50\%&Undervolt+RFT&93.04&100/48&87/53&100/57&86/58&58/27&100/73&95/69&53/5&100/64&92/60\\

&Undervolt+CDFT&89.89&100/49&92/51&100/61&94/59&76/38&100/65&96/66&70/10&100/63&96/62\\
\hline

\end{tabular}
\endminipage
}

\end{table*}



















\subsection{Sensitivity to Error Profile }

We investigate the effect of different error profiles on the effectiveness of our design. We model single bit errors as well as multi-bit errors covering 25\% and 50\% of the bits in a word. We expect undervolting to produce multi-bit errors, and therefore use 50\% bit flips in our main results. For completeness, however, we include results for single bit errors as well as 25\% bit errors. 


\begin{figure}[h]
\centering
     \includegraphics[width=\linewidth]{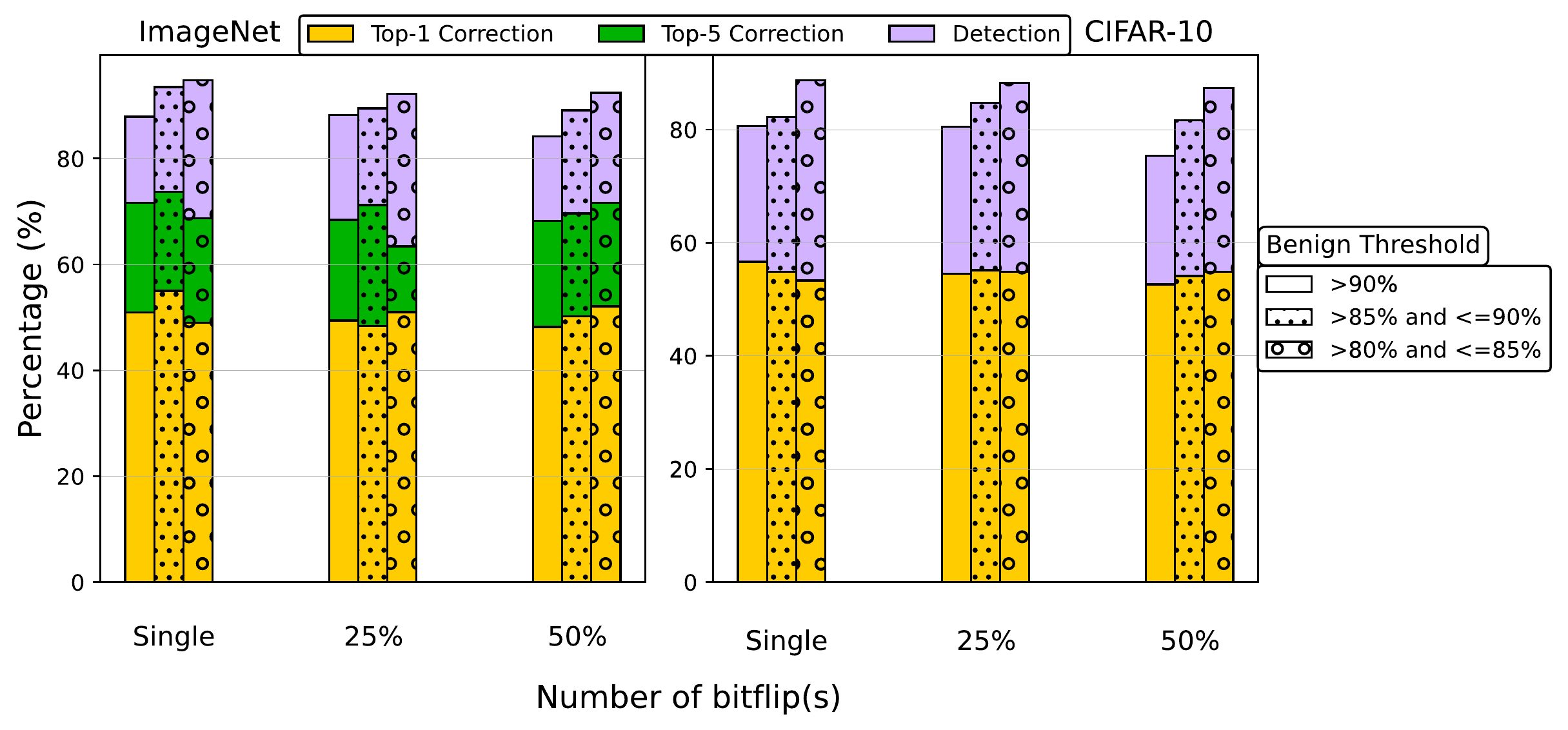}
  \caption{\small Adversarial detection and correction for different error profiles at multiple benign true positive rate thresholds.}\label{fig:BitComparision_VGG_Densenet}
\end{figure}

Figure \ref{fig:BitComparision_VGG_Densenet} shows the average detection and correction rates for the three error profiles. We show results for three different thresholds for benign accuracy:  $>$80\%, $>$85\% and $>$90\%. For each experiment we use the lowest error rate that is sufficient to ensure the benign accuracy threshold is met. We can see that the correction and detection rates are remarkably consistent across different error profiles for both ImageNet and CIFAR data sets. The detection rates varies by at most 4\% between different rates of bit flips and when considering a benign threshold TPR of greater than 90\% for both the datasets. 

Table \ref{tab:imagenet_attack_succ} shows a summary of the detection and correction rates across all attacks and error profiles we evaluate for the ImageNet dataset. For each error profile, we use an error rate that is sufficient to achieve a $>$85\% benign TPR for the main defense (Undervolt+CDFT).

\subsection{Comparison with Prior Work}

\begin{figure}[h]
\centering
     \includegraphics[width=\linewidth]{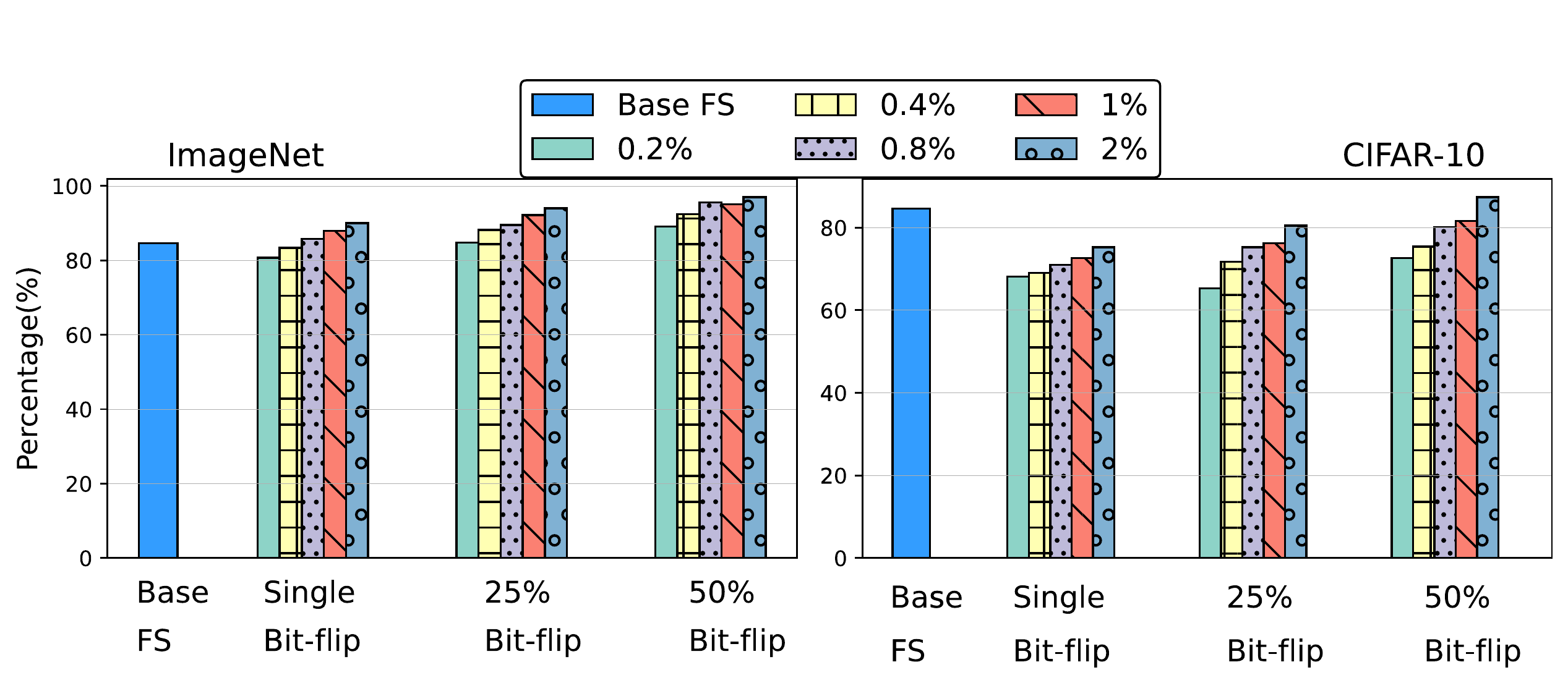}
  \caption{\small Adversarial detection compared with Feature Squeezing.}\label{fig:FSCompare}
\end{figure}

We compare our defense with Feature Squeezing (FS) \cite{xu2018feature}, a defense that manipulates, or "squeezes", input images (for example by reducing the color depth and smoothing to reduce the variation among pixels). If the low-resolution image produces substantially different classification outputs than the original image, the input is likely to be adversarial. FS combines multiple squeezers for detection resulting in higher performance overhead than our proposed technique. We applied FS to our set of adversarial images for both ImageNet and CIFAR-10. Figure \ref{fig:FSCompare} compares the average adversarial detection rate of FS and our defense for different error rates. We observe that our design compares favorably to FS, even for low error rates. As the error rate increases, our technique performs better, reaching a 97\% adversarial detection rate.


We also observe that FS can complement our defense to achieve higher detection rate. Table-\ref{tab:fs-undervolt-summary} shows detection rates for a FS variant (using a 2x2 median filter smoothing) compared to our main design (Undervolting+CDFT) and to our design that includes FS (Undervolting+CDFT+FS). 
We can see that the combined design achieves the highest detection rate for both datasets, demonstrating that undervolting can potentially complement other defenses. However, we do note that the benign TPR decreases slightly for the combined technique. Additional tuning of the combined design might be needed, but was beyond the scope of this work.

\begin{table}
\caption{\small Average adversarial detection rate using a fixed benign TPR threshold for ImageNet/CIFAR-10.}
\label{tab:fs-undervolt-summary}
\small
\centering
\scalebox{1}{
\begin{tabular}{|c||c|c|}
\hline

Technique&
Benign TPR(\%)&
Detection(\%)  \\

\hline
FS(2x2)&93.80/93.20&79.40/85.40\\
UV+CDFT&92.93/93.50&85.20/81.25\\
UV+CDFT+FS(2x2)&90.40/91.70&91.60/89.50\\
\hline

\end{tabular}
}

\end{table}

\subsection{Resilience to Defense-Aware Attacks}

We evaluate our solution against attackers that are aware of the details of our defense. We examine two such attacks: 

\subsubsection{Attack Trained Under Random Noise}

In the first attack we assume the adversary is aware of the error-based defense, but has no access to the victim's hardware (Figure \ref{fig:adaptive} \circled{1}) and does not know the error distribution profile of the victim's device. The attack adds random noise into the model during adversarial construction to attempt to circumvent our defense.

\subsubsection{Attack Trained on Device}

\begin{figure}[h]
\centering
     \includegraphics[width=\linewidth]{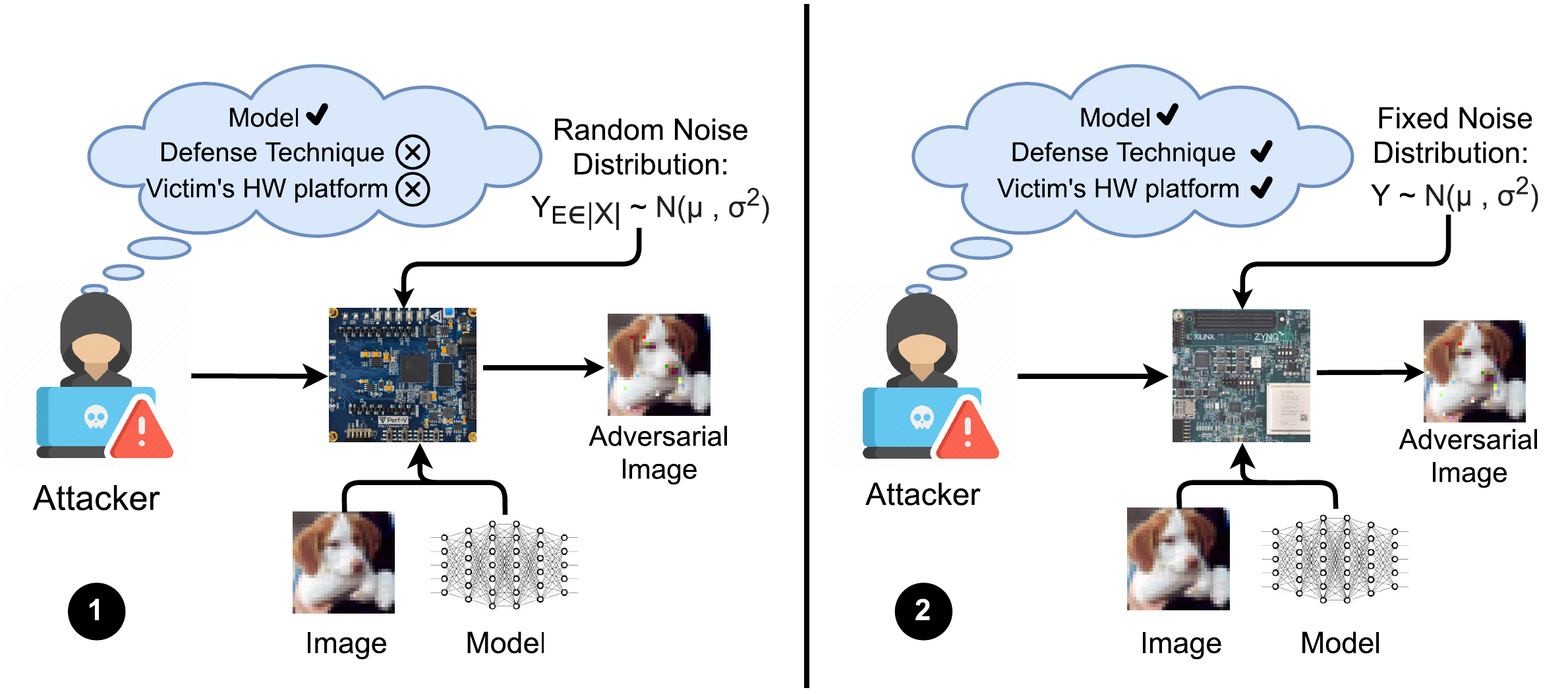}
  \caption{\small Defense aware attacks scenarios: (1) trained with random noise and (2) trained on device that includes the defense.}\label{fig:adaptive}
\end{figure}

In the second scenario, we assume an attacker with full control of a hardware platform that includes our defense, but is not the device under attack. Control of the device allows the attacker to lean the error distribution of the platform, the model parameters, and can use it indefinitely to create adversarial attacks directly, as shown in Figure \ref{fig:adaptive} \circled{2}. The attack is constructed by training on a device with a pre-determined error profile. This is to determine if an attacker controlling one device can successfully attack a different device, with different error characteristics. 

 
We implement both defense-aware attacks on top of the $CW_{L2}$ attack and use them to generate adversarials based on the ImageNet dataset. All generated adversarials have a baseline success rate of 100\%. We re-evaluate these adversarials on a different simulated device with a different error profile. Table \ref{tab:attackwbtab} shows the detection rate for our design for different error rates. We can see that the detection rate for both attacks is very high. This shows that the unique error characteristics of each device ensure a robust defense against defense-aware adversarials. 


         

\begin{table}
\caption{\small Detection rates of defense-aware attacks at different error rates.}
\label{tab:attackwbtab}
\small
\centering
\scalebox{1}{
\begin{tabular}{|c||c|c|}
\hline

\multirow{2}{*}{Error Rate}&
\multicolumn{2}{c|}{Detection Rate}\\
\cline{2-3}
&
\multicolumn{1}{m{3cm}|}{Random Noise Attack}&
\multicolumn{1}{m{3cm}|}{Device Trained Attack} \\

\hline
0.2\%&88\%&81\%\\
0.4\%&90\%&88\%\\
0.8\%&96\%&95\%\\
1\%&99\%&98\%\\
2\%&100\%&100\%\\
\hline

\end{tabular}
}
\end{table}

\subsection{Energy Impact}

Finally, we evaluate the energy impact of the proposed defense, using measurements of power consumption and execution time performed on our FPGA-based implementation of ChaiDNN running VGG16. We measure FPGA power while running the CHaiDNN model, using the external voltage controller. Table \ref{tab:chaienergy} shows inference energy for our defense relative to the unprotected baseline for both correction and detection. Due to the substantial power reduction achieved by undervolting and given that our correction method has no performance overhead, our correction defense achieves about 30\% energy savings. Detection requires two sequential inference passes, but due to undervolting this comes at a 70\% increase in energy cost instead of 2$\times$.



    
         

\begin{table}[htb]
\caption{\small Relative energy for adversarial correction and detection with critical voltage in ZCU104 running CHaiDNN.}
\label{tab:chaienergy}
\small
\centering
\scalebox{1}{
\begin{tabular}{|c||c|c|}
\hline

\multirow{2}{*}{Critical Voltage}&
\multicolumn{2}{c|}{Relative Energy}\\
\cline{2-3}
&
\multicolumn{1}{m{1.5cm}|}{Correction}&
\multicolumn{1}{m{1.5cm}|}{Detection} \\

\hline
0.660V&0.716&1.716\\
0.656V&0.714&1.714\\
0.652V&0.670&1.670\\
\hline

\end{tabular}
}
\end{table}
\section{Conclusion}
\label{sec:conclusion}

In conclusion, this paper proposes a novel defense against adversarial ML attacks that relies on undervolting . We show that the technique is highly effective, with $>90$\% detection rate of adversarial inputs for a variety of state-of-the-art attacks. The defense is also energy efficient, due to the power savings from low voltage operation. We also show how the randomness introduced by the unique device characteristics makes our approach robust against defense-aware attacks. We show that the defense is not device specific and can work, with different degrees of effectiveness, depending on the errors produced by undervolting on different devices. 

\bibliographystyle{IEEEtranS}
\bibliography{main}

\end{document}